\documentclass[a4paper, 11pt]{article} 
\usepackage{amsmath,amssymb,enumerate, algorithm, algorithmic, subfig, xspace, wrapfig, graphicx, caption,multicol, array, bbm}
\usepackage{jmlr2e}
\usepackage{url}
\usepackage{etoolbox}
\usepackage{natbib}
\usepackage{hyperref} 
\usepackage{placeins}
\usepackage{tikz,enumerate}
\usepackage[framemethod=TikZ]{mdframed}
\renewcommand{\baselinestretch}{1.05}
\usetikzlibrary{arrows}
\newdimen\arrowsize
\pgfarrowsdeclare{arcsq}{arcsq}
{
  \arrowsize=0.2pt
  \advance\arrowsize by .5\pgflinewidth
  \pgfarrowsleftextend{-4\arrowsize-.5\pgflinewidth}
  \pgfarrowsrightextend{.5\pgflinewidth}
}
{
  \arrowsize=1.5pt
  \advance\arrowsize by .5\pgflinewidth
  \pgfsetdash{}{0pt} 
  \pgfsetroundjoin   
  \pgfsetroundcap    
  \pgfpathmoveto{\pgfpoint{0\arrowsize}{0\arrowsize}}
  \pgfpatharc{-90}{-140}{4\arrowsize}
  \pgfusepathqstroke
  \pgfpathmoveto{\pgfpointorigin}
  \pgfpatharc{90}{140}{4\arrowsize}
  \pgfusepathqstroke
}
\usepackage{fullpage}

\definecolor{mauve}{rgb}{0.88, 0.69, 1.0}
\definecolor{ferngreen}{rgb}{0.31, 0.47, 0.26}

\usepackage{listings}
\renewcommand{\algorithmicrequire}{\textbf{Input:}}
\renewcommand{\algorithmicensure}{\textbf{Output:}}

\def\abovestrut#1{\rule[0in]{0in}{#1}\ignorespaces}
\def\belowstrut#1{\rule[-#1]{0in}{#1}\ignorespaces}

\def\abovespace{\abovestrut{0.15in}}

\def\belowspace{\belowstrut{0.075in}}

\newcommand{\independent}{\mbox{${}\perp\mkern-11mu\perp{}$}}
\newcommand{\notindependent}{\mbox{${}\not\!\perp\mkern-11mu\perp{}$}}

\newcommand{\hypnonlinICP}{$H_{0,S}$\xspace}

\newcommand{\Xs}{X_S}
\newcommand{\causalvars}{X_{S^*}}
\newcommand{\causalset}{S^*}

\newcommand{\pval}{pv}
\newcommand{\blocklen}{l_{b}}

\newcommand{\envvar}{E}

\newcommand{\E}{{\mathcal{E}}}
\newcommand{\given}{\,|\,}
\newcommand{\mean}{{\mathbf{E}}}

\newcommand{\ACE}{{\mathit{ACE}}}

\usepackage[framemethod=TikZ]{mdframed}
\usepackage{paralist}
\usetikzlibrary{arrows}

\title{Invariant Causal Prediction for Nonlinear Models}

\author{\name Christina Heinze-Deml \email \url{heinzedeml@stat.math.ethz.ch}\\ 
\addr Seminar f\"ur Statistik, ETH Zurich, Switzerland
\AND  
   \name Jonas Peters \email \url{jonas.peters@math.ku.dk}\\
\addr University of Copenhagen, Denmark
\AND  
   \name Nicolai Meinshausen \email \url{meinshausen@stat.math.ethz.ch}\\
\addr Seminar f\"ur Statistik, ETH Zurich, Switzerland}

\editor{}

\begin{document}

\maketitle

\begin{abstract}
An important problem in many domains is to predict how a system will
respond to interventions. This task is inherently linked to estimating
the system's underlying causal structure. To this end, Invariant
Causal Prediction (ICP) 
\citep{PetBuhMei15} has been proposed which learns a causal model exploiting the invariance of causal relations using data from  different environments. When considering linear models, the implementation of ICP is relatively straightforward. However, the nonlinear case is more challenging due to the difficulty of 
performing nonparametric tests for conditional independence.  

In this work, we present and evaluate an array of methods for nonlinear and nonparametric versions of ICP  for learning the 
causal  parents of given target variables. 
We find that an approach which first fits a nonlinear model with data pooled over all environments and then tests for differences between the residual distributions across environments is quite robust across a large variety of simulation settings. 
We call this procedure ``invariant residual distribution test''.
In general, we observe that the performance of all approaches is critically dependent 
on the true (unknown) causal structure and it becomes challenging to achieve high power if the parental set  includes more than two variables. 

As a real-world example, we consider fertility rate modeling which is central to world population projections. We explore predicting the effect of hypothetical interventions using the accepted models from nonlinear ICP. The results reaffirm the previously observed  central causal role of child mortality rates.

\end{abstract}

\section{Introduction}
Invariance based causal discovery \citep{PetBuhMei15} relies on the observation that the conditional distribution of the target variable $Y$ given its direct causes remains invariant if we intervene on variables other than $Y$. While the proposed methodology in \cite{PetBuhMei15} focuses on linear models, we extend Invariant Causal Prediction to nonlinear settings. We first introduce the considered structural causal models in Section~\ref{subsec:scm} and review related approaches to causal discovery in Section~\ref{subsec:causal_dis}. The invariance approach to causal discovery from \cite{PetBuhMei15} is briefly summarized in Section~\ref{subsec:invariance_discovery} and we outline our contribution in Section~\ref{subsec:contribution}. In Section~\ref{subsec:fetility_rate_modelling} we introduce the problem of fertility rate modeling which we consider as a real-world example throughout this work.

\subsection{Structural causal models}\label{subsec:scm}
Assume an underlying structural causal model (also called structural equation model) \citep[e.g.][]{Pearl2009}
\begin{eqnarray*}
Z_1  &\;\leftarrow \;&  g_1(Z_{\text{pa}_1}) + \eta_1,\\
Z_2  &\;\leftarrow \;&  g_2(Z_{\text{pa}_2}) + \eta_2,\\
&\vdots&  \\
Z_q  &\;\leftarrow \;&  g_q(Z_{\text{pa}_q}) + \eta_q,
\end{eqnarray*}
for which the functions $g_{k}$, $k=1, \ldots, q$,  as well as the parents
$\text{pa}_k\subseteq\{1,\ldots,q\}\setminus \{k\}$ of each variable
are unknown. 
Here, we have used the notation $Z_S=(Z_{i_1},\ldots,Z_{i_s})$ for
any set $S=\{i_1,\ldots,i_s\} \subseteq\{1,\ldots,q\}$. 
We assume the corresponding directed graph to be acyclic. We further require the noise variables $\eta_1, \ldots, \eta_q$ to be jointly independent and to have zero mean, i.e.\ we assume that there are no hidden variables.

Due to its acyclic structure, it is apparent that such a structural causal model induces a joint distribution $P$ over the observed random variables. 
Interventions on the system are usually modeled by replacing some of the structural assignments 
\citep[e.g.][]{Pearl2009}.
If one intervenes on variable $Z_3$, for example, and sets it to the value $5$, 
the system again induces a distribution over $Z_1, \ldots, Z_q$, that 
we denote by $P(\cdot | do(Z_3 \leftarrow 5))$. 
It is  usually different from the observational distribution $P$. 
We make no counterfactual assumptions here: we assume a new
realization  $\eta$ is drawn from the noise distribution as soon as we make
an intervention.\footnote{The new  realization of $\eta$ under an
  intervention and the  realization
  under observational data can be assumed to be independent. However,
  such an assumption is untestable since we can never observe
  realizations under  different interventions simultaneously and we do not make statements or assumptions about the joint distribution of observational and interventional settings.  
}  

\subsection{Causal discovery}\label{subsec:causal_dis}
In causal discovery (also called structure learning) one tries to reconstruct the structural causal model or its graphical representation from its joint distribution \citep[e.g.][]{Pearl2009, Spirtes2000, Peters2017, Chickering2002, Peters2014, Heckerman1997, hapb14}.

Existing methods for causal structure learning can be categorized along a number of dimensions, such as (i) using purely observational data vs.\ using a combination of interventional and observational data; (ii) score-based vs.\ constraint-based vs.\ ``other'' methods; (iii) allowing vs.\ precluding the existence of hidden confounders; (iv) requiring vs.\ not requiring faithfulness;\footnote{
A distribution satisfies faithfulness and the global Markov condition with respect to a graph $G$ if the following statement holds for all disjoint sets $A$, $B$, and $C$ of variables: $A$ is independent of $B$, given $C$, if and only if $A$ is $d$-separated (in $G$) from $B$, given $C$. 
The concept of $d$-separation \citep{Pearl1985, Pearl1988book} is defined in \citet[][Def.~6.1]{Peters2017}, for example.
} (v) type of object that the method estimates. Moreover, different methods vary by additional assumptions they require.
In the following, we give brief descriptions of the most common methods for causal structure learning\footnote{Also see \cite{HeinzeDeml2018} for a review and empirical comparison of recently proposed causal structure learning algorithms.}. 

The PC algorithm \citep{Spirtes2000} uses observational data only and estimates the Markov equivalence class of the underlying graph structure, based on (conditional) independence tests under a faithfulness assumption. The presence of hidden confounders is not allowed. Based on the PC algorithm, the IDA algorithm \citep{Maathuis2009} computes bounds on the identifiable causal effects. 

The FCI algorithm is a modification of the PC algorithm. It also relies on purely observational data while it allows for hidden confounders. 
The output of FCI is a partial ancestral graph (PAG), i.e.\ it estimates the Markov equivalence class of the underlying maximal ancestral graph (MAG). 
Faster versions, RFCI and FCI+, were proposed by \citet{Colombo2012} and 
\citet{Claassen2013}, respectively.

The PC, FCI, RFCI and FCI+ algorithms are formulated such that they allow for an independence oracle 
that indicates whether a particular (conditional) independence holds
in the distribution. These algorithms are  typically  applied in the linear Gaussian setting where testing for conditional independence reduces to testing for vanishing partial correlation.

One of the most commonly known score-based methods is greedy equivalence search (GES). Using observational data, it greedily searches over equivalence classes of directed acyclic graphs for the best scoring graph (all graphs within the equivalence class receive the same score) where the score is given by the Bayesian information criterion, for example. Thus, GES is based on an assumed parametric model such as linear Gaussian structural equations or multinomial distributions. The output of GES is the estimated Markov equivalence class of the underlying graph structure. 
\citet{Heckerman1997}
describe a score-based method with a Bayesian score.

Greedy interventional equivalence search (GIES) extends GES to operate on a combination of interventional and observational data. The targets of the interventions need to be known and the output of GIES is the estimated interventional Markov equivalence class. The latter is typically smaller than the Markov equivalence class obtained when using purely observational data. 

Another group of methods makes restrictive assumptions which allows for obtaining full identifiability. Such assumptions include non-Gaussianity \citep{Shimizu2006} or equal variances \citep{Peters2012} of the errors or non-linearity of the structural equations in additive noise models \citep{Hoyer2008, Peters2014}.

Instead of trying to infer the whole graph, 
we are
here interested in settings, where 
there is a target variable $Y$ of special interest. 
The goal is to infer both the parental set $S^*$ for the target
variable $Y$ and confidence bands for the causal effects.

\subsection{Invariance based causal discovery}\label{subsec:invariance_discovery}
This work builds on the method of Invariant Causal Prediction (ICP) \citep{PetBuhMei15} and extends it in several ways. 
The method's key observation is 
that 
the conditional distribution of the target variable $Y$ given its
direct causes remains invariant if we intervene on variables other
than $Y$. This follows from  an assumption sometimes called autonomy or modularity \citep{Haavelmo1944, Aldrich1989, Hoover1990, Pearl2009, Schoelkopf2012}.
In a linear setting, this implies, for example, that regressing $Y$ on its direct causes yields the same regression coefficients in each environment, provided we have an infinite amount of data. 
In a nonlinear setting, this can be generalized to a conditional independence
between an index variable indicating the interventional setting and $Y$, given $X$; see~\eqref{eq:H0S}.
The method of ICP  assumes that we are given data from several environments. 
It searches for sets of covariates, for which the above property of invariance cannot be rejected. The method then outputs the intersection of all such sets, which can be shown to be a subset of the true set with high probability, see Section~\ref{subsec:nonlinICP} and Algorithm~\ref{alg:icp_generic} in Appendix~\ref{supp:sec:condtest_details} for more details. Such a 
coverage guarantee is highly desirable, especially in causal discovery, where information about ground truth is often sparse.

In many real life scenarios, however, relationships are not linear and the above procedure can fail: The true set does not necessarily yield an invariant model and the method may lose its coverage guarantee, see Example~\ref{ex:counterlinear}. 
Furthermore, environments may not come as a categorical variable but as a continuous variable instead. In this work, we extend the concept of ICP to nonlinear settings and continuous environments. The following paragraph summarizes our contributions.

\subsection{Contribution}\label{subsec:contribution}
Our contributions are fivefold.

\paragraph{Conditional independence tests.}
We extend the method of ICP to nonlinear settings by considering
conditional independence tests. We  discuss in Section~\ref{subsec:condtest} and in more length in Appendix~\ref{supp:sec:condtest_details} several possible nonlinear and nonparametric tests for conditional independence
  of the type~\eqref{eq:H0S} and propose alternatives. 
  There has been some progress
  towards nonparametric independence tests \citep{bergsma2014consistent, hoeffding1948, blum1961, Renyi1959, szekely2007, Zhang11}. However,  in the general nonparametric
  case, no known 
non-trivial
  test of conditional independence has
  (even asymptotically)
  a type I error rate less than the pre-specified significance level. 
  This stresses the importance of empirical evaluation of conditional independence tests.

\paragraph{Defining sets.} 
We  discuss in Section~\ref{subsec:defining} cases of poor identifiability of the causal
  parents. If there are highly correlated variables in the dataset, we
  might get an empty estimator if we follow the approach proposed
  in~\citep{PetBuhMei15}.   We can, however, extract more information via defining
  sets. The results are to some extent comparable to similar issues
  arising in multiple testing \citep{goeman2011multiple}. For example,
  if  we know that the parental set of  a variable $Y$ is either
  $S=\{1,3\}$  or $S=\{2,3\}$, we know that $\{3\}$ has to be a parent
  of $Y$. Yet we also want to explore the information that one
  variable out of the set $\{1,2\}$ also has to be causal for the
  target variable~$Y$,
  even if we do not know which one out of the two.

\paragraph{Confidence bands for causal effects.} 
Beyond identifying the causal parents, we can provide
  nonparametric or nonlinear confidence bands for the strength of the
  causal effects, as shown in Section~\ref{subsec:bands}.

\paragraph{Prediction under interventions.} 
Using the accepted models from nonlinear ICP, we are able to forecast the average causal effect of external interventions. We will discuss
  this at hand of examples in Section~\ref{subsec:prediction}.

\paragraph{Software.} \textsf{R} \citep{R-language} code for nonlinear ICP is provided in the package \texttt{nonlinearICP}. The proposed conditional independence tests are part of the package \texttt{CondIndTests}. Both packages are available from CRAN.

\subsection{Fertility rate modeling}\label{subsec:fetility_rate_modelling}
At the hand of the example of fertility rate modeling, we shall explore how to
exploit the invariance of causal models for causal discovery in the
nonlinear case.

Developing countries have a significantly higher fertility rate
compared to Western countries. The fertility rate can be predicted
well from covariates such as `infant mortality rate' or `GDP per
capita'. Classical prediction models, however, do not answer whether
an active intervention on some of the covariates leads to a change in
the fertility rate. This can only be answered by exploiting causal
knowledge of the system. 

Traditionally, in statistics the methods for establishing causal relations rely on carefully designed randomized studies. 
Often, however, 
such experiments cannot be performed. For instance, factors like `infant mortality rate' are
highly complex and cannot be changed in isolation.
We may still be interested in the effect of a policy that aims at
reducing the infant mortality rate but this policy cannot be randomly
assigned to different groups of people within a country. 

There is a large body of work that
is trying to explain changes in fertility; for an interesting
overview of different theories see \citet{hirschman1994fertility} and
the more recent \citet{huinink2015explaining}. There is not a single
established theory for changes in fertility and we should 
clarify in the beginning that all models we will be using will have
shortcomings, especially the shortcoming that we might not have
observed all relevant variables. We would nevertheless like to take the
fertility data as an example to establish a methodology that allows
data-driven answers; discussing potential
shortfalls of the model is encouraged and could be beneficial in
further phrasing the right follow-up questions and collecting perhaps
more suitable data.

An interesting starting point for us was the work of 
\citet{raftery1995demand} and very helpful discussions with co-author Adrian
Raftery. That work tries to distinguish between two different explanatory
models for a decline in fertility in Iran. One model argues that
economic growth is mainly responsible; another argues that
transmission of new ideas is the primary factor (ideation
theory). What allows a distinction between these models is that
massive economic growth started in 1955 whereas ideational changes
occurred mostly 1967 and later. Since the fertility began to drop
measurably already in 1959, the demand theory seems more plausible and
the authors conclude that reduced child mortality is the key
explanatory variable for the reduction in fertility (responsible for at least a quarter of the
reduction). 

Note the way we decide between  two potential  causal theories for a decline in fertility: if a causal model is valid, it has
to be able to explain the decline consistently. In particular, the
predictions of the model have to be valid for all
time-periods, including the time of 1959 with the onset of the
fertility decline. The ideation theory wrongly places the onset of
fertility decline later and is thus dismissed as less plausible.

The invariance approach of \citet{PetBuhMei15} we follow here for linear models has a similar
basic idea: a causal model has to work consistently. In our case, we choose
geographic location instead of time for the example and demand that a
causal model has to work consistently across geographic locations or
continents. We collect all potential models that show this invariance
and know that \emph{if} the underlying assumption of causal
sufficiency is true and we have observed all important causal
variables \emph{then} the causal model will be in the set of retained
models. Clearly, there is room for a healthy and interesting debate to what extent the causal
sufficiency assumption is violated in the example. It has been argued, however, that missing variables do not allow for any invariant model, which renders the method to remain conservative \citep[][Prop.~5]{PetBuhMei15}.

We establish a framework
for causal discovery in nonlinear models. Incidentally, the approach also identifies reduced child
mortality as one of  key explanatory variables for a decline in
fertility.

\section{Nonlinear Invariant Causal Prediction}

We first extend the approach of~\citep{PetBuhMei15} to nonlinear
models, before discussing defining sets, nonparametric confidence
bands and prediction under interventions.

\subsection{Invariance approach for causal discovery} \label{subsec:nonlinICP}
\cite{PetBuhMei15} proposed an invariance approach in the context of
linear models. We describe the approach here in 
a notationally slightly different way that will simplify statements and
results in the nonlinear case and allow for more general
applications. 
Assume that we are given a structural causal model (SCM) over
variables $(Y,X,E)$, where $Y$ is the target variable, $X$ the
predictors and $E$ so-called environmental variables. 
\begin{definition}[Environmental variables] \label{def:env}
We know or assume that the variables  $E$ are neither descendants nor
parents of $Y$ in the causal DAG of $(Y,X,E)$.
If this is the case, we call $E$ environmental variables.
\end{definition}
In \cite{PetBuhMei15}, the environmental variables were 
given and non-random.
Note that the definition above treats the variables as random but we
can in practice condition on the observed values of $E$. The
definition above excludes the 
possibility that there 
is a direct causal connection between one of the variables in $\envvar$
and $Y$. We will talk in the following about the triple of random
variables $(Y,X,E)$, where the variable $X$ of predictor variables is
indexed by  $X_1, \ldots, X_p$. With a slight abuse of notation, we
 let $S^*\subseteq\{1,\ldots,p\}$ be the indices of $X$ that are
 causal parents $\text{pa}_Y$ of $Y$.
Thus, the structural equation for $Y$ can be written as
\begin{equation}\label{eq:Y}   
Y \;\leftarrow\; f( X_{S^*}) + \varepsilon ,\end{equation}
where 
$f:\mathbb{R}^{\vert S^*\vert} \to \mathcal{Y}$.
We let $\mathcal{F}$ be the function class of $f$ and let
$\mathcal{F}_S$ be the subclass of functions that depend only on the set
$S\subseteq\{1,\ldots,p\}$ of variables.
With this notation we have $f\in \mathcal{F}_{S^*}$.

 The assumption of no direct effect of $E$ on $Y$ is analogous to the typical assumptions about
instrumental variables \citep{angrist1996identification,imbens2014instrumental}. See Section~5 in \cite{PetBuhMei15}  for a longer
discussion on the relation between environmental variables and
instrumental variables. The two main distinctions between
environmental and instrumental variables are as follows. First, we do not need to test for
the ``weakness'' of instrumental/environmental variables since we do
not assume that there is a causal effect from $E$ on the variables in $X$. Second,  the
approaches are used in different contexts. With instrumental
variables, we assume the graph structure to be known typically and
want to estimate the strength of the causal connections, whereas the
emphasis is here on both causal discovery (what are the parents of a
target?) and then also inference for the strength of causal
effects. With a single environmental variable, we can identify in some
cases multiple causal effects whereas the number of instrumental
variables needs to match or exceed the number of variables in
instrumental variable regression.
The instrumental variable approach, on the other hand, can correct for unobserved confounders between the parents and the target variable if their influence is linear, for example. In these cases, our approach could remain uninformative \citep[][Proposition~5]{PetBuhMei15}. 

\begin{figure}[ht]
\begin{center}
\begin{tikzpicture}[scale=1, line width=0.5pt, minimum size=0.58cm, inner sep=0.3mm, shorten >=1pt, shorten <=1pt]
    \normalsize
    \draw (2,2) node(c) [rectangle,minimum size=6mm,rounded corners=2mm, draw] {\small Continent};
    \draw (-1,0) node(x1) [rectangle,minimum size=6mm,rounded corners=2mm, draw]   {\small Education};
    \draw (-0.5,2) node(x3) [rectangle,minimum size=6mm,rounded corners=2mm, draw]   {\small \phantom{kk}GDP\phantom{kk}};
    \draw (1.5,0) node(y) [rectangle,minimum size=6mm,rounded corners=2mm, draw] {\small \phantom{kk}TFR\phantom{kk}};
    \draw (2,-1.7) node(x2) [rectangle,minimum size=6mm,rounded corners=2mm, draw] {\small \phantom{kk}IMR\phantom{kk}};
    \draw [-arcsq] (c) to (x1);
    \draw [-arcsq] (c) to (x3);
    \draw [-arcsq] (x3) to (x1);
    \draw [-arcsq] (x1) to  (x2);
     \draw [-arcsq] (y) to (x2);
    \draw [-arcsq] (x1) to  (y);
   \end{tikzpicture}
     \hfill
\begin{tikzpicture}[scale=1, line width=0.5pt, minimum size=0.58cm, inner sep=0.6mm, shorten >=1pt, shorten <=1pt]
    \normalsize
    \draw (2,2) node(c) [rectangle,minimum size=6mm,rounded corners=2mm, draw] {\small Continent};
    \draw (-1,0) node(x1) [rectangle,minimum size=6mm,rounded corners=2mm, draw]   {\small Education};
    \draw (-0.5,2) node(x3) [rectangle,minimum size=6mm,rounded corners=2mm, draw]   {\small \phantom{kk}GDP\phantom{kk}};
    \draw (1.5,0) node(y) [rectangle,minimum size=6mm,rounded corners=2mm, draw] {\small \phantom{kk}TFR\phantom{kk}};
    \draw (2,-1.7) node(x2) [rectangle,minimum size=6mm,rounded corners=2mm, draw] {\small \phantom{kk}IMR\phantom{kk}};
    \draw [-arcsq] (c) to[bend left] (x2);
    \draw [-arcsq] (c) to (x3);
    \draw [-arcsq] (x3) to (x1);
    \draw [-arcsq] (y) to  (x1);
     \draw [-arcsq] (x2) to (y);
    \draw [-arcsq] (x3) to  (y);
   \end{tikzpicture}
   \hfill
\begin{tikzpicture}[scale=1, line width=0.5pt, minimum size=0.58cm, inner sep=0.3mm, shorten >=1pt, shorten <=1pt]
    \normalsize
    \draw (2,2) node(c) [rectangle,minimum size=6mm,rounded corners=2mm, draw] {\small Continent};
    \draw (-1,0) node(x1) [rectangle,minimum size=6mm,rounded corners=2mm, draw]   {\small Education};
    \draw (-0.5,2) node(x3) [rectangle,minimum size=6mm,rounded corners=2mm, draw]   {\small \phantom{kk}GDP\phantom{kk}};
    \draw (1.5,0) node(y) [rectangle,minimum size=6mm,rounded corners=2mm, draw] {\small \phantom{kk}TFR\phantom{kk}};
    \draw (2,-1.7) node(x2) [rectangle,minimum size=6mm,rounded corners=2mm, draw] {\small \phantom{kk}IMR\phantom{kk}};
    \draw [-arcsq] (c) to[bend left] (x2);
    \draw [-arcsq] (c) to (x3);
    \draw [-arcsq] (x3) to (x1);
    \draw [-arcsq] (y) to  (x1);
    \draw [-arcsq] (c) to  (y);
     \draw [-arcsq] (x2) to (y);
    \draw [-arcsq] (x3) to  (y);
   \end{tikzpicture}
\caption{\small Three candidates for a causal DAG with target total fertility 
rate (TFR) and four potential causal predictor variables. We would like 
to infer the parents of TFR in the true causal graph. We use the 
continent as the environment variable  $\envvar$. If the true DAG was one of the 
two graphs on the left, the environmental variable would have no direct 
influence on the target variable TFR and `Continent' would be a valid 
environmental variable, see Definition~\ref{def:env}.
 }
 \label{fig:fert_graphs}
\end{center}
\end{figure}
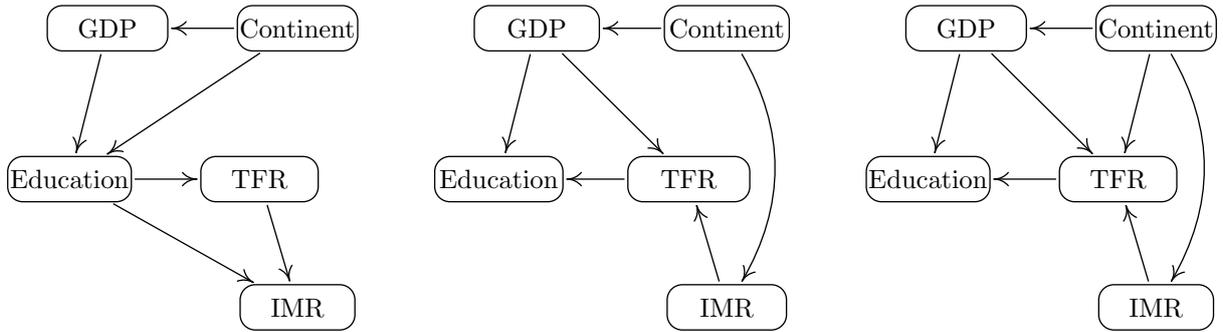

\paragraph{Example (Fertility data).}
In this work, we analyze a data set provided by the \cite{fertility_data}. 
Here, 
$Y,X$ and $\envvar$ correspond to the following quantities:
\begin{enumerate}[(a)]
\item $Y\in \mathbb{R}$ is the total fertility rate (TFR) in a country in a
  given year,
\item $X\in \mathbb{R}^9$ are potential causal predictor
  variables for TFR: 
  \begin{compactitem}
  	\item[--] IMR -- infant mortality rate
  	\item[--] Q5 -- under-five mortality rate
  	\item[--] Education expenditure (\% of GNI)
  	\item[--] Exports of goods and services (\% of GDP)
  	\item[--] GDP per capita (constant 2005 US\$)
  	\item[--] GDP per capita growth (annual \%)
  	\item[--] Imports of goods and services (\% of GDP)
  	\item[--] Primary education (\% female)
  	\item[--] Urban population (\% of total)
  \end{compactitem}
\item $\envvar \in \{ C_1, C_2, C_3, C_4, C_5, C_6\}$
is the continent of the country, divided
  into the categories Africa, Asia, Europe, North and South America
  and Oceania. If viewed as a random variable (which one can argue
  about), the assumption is that the continent is not a descendant of
  the fertility rate, which seems plausible.
  For an environmental variable, the additional
  assumption is that the TFR in a country is only indirectly
  (that is via one of the other variables) influenced by which
  continent it is situated on (cf.\ Figure~\ref{fig:fert_graphs}).
\end{enumerate}
Clearly, the choices above are debatable. We might for example also
want to include some ideation-based variables in $X$ (which are harder
to measure, though) and also take different environmental variables $E$
such as time instead of geographic location. We could even allow for
additive effects of the environmental variable on the outcome of
interest (such as a constant  offset for each continent) but we do not
touch this debate much more here as we are  primarily interested in the
methodological development.

The basic yet central insight underlying the invariance approach  is
the fact that for the true causal parental set $S^* := \text{pa}_Y$
we have the following conditional independence relation
under Definition~\ref{def:env} of environmental variables:
\begin{equation}\label{eq:ind_relation}
Y \independent \envvar \; \vert \; \causalvars. 
\end{equation}
This follows directly from the local Markov condition \citep[e.g.][]{Lauritzen1996}.
The goal is to find $\causalset$  by exploiting the above relation~\eqref{eq:ind_relation}.
Suppose we have a test for the null hypothesis 
\begin{equation}\label{eq:H0S} H_{0,S}: \qquad Y \independent \envvar  \; \vert \; X_S.\end{equation}
It was then proposed in \cite{PetBuhMei15} to define an estimate
$\hat{S}$ for the parental set $S^*$ by setting
\begin{equation}\label{eq:Shat}  \hat{S} := \bigcap_{S:H_{0,S} \text{ not rejected}} S .
\end{equation}
Here, the intersection runs over all sets $S$, s.t.\ $E \cap S = \emptyset$.
If the index set is empty, i.e.\ $H_{0,S}$ is rejected for all 
sets $S$,
we define $\hat{S}$ to be the empty set.
If we can test~\eqref{eq:H0S} with the correct type I error rate in the sense that 
\begin{equation} \label{eq:level} P\big( H_{0,S^*} \mbox{ is rejected at level }  \alpha \big) \leq \alpha,
\end{equation}
then we have as immediate consequence the desired statement
\[ P\big( \hat{S} \subseteq S^* \big) \;\ge\; P\big( H_{0,S^*} \mbox{ accepted} \big) \;\ge\; 1-\alpha .\]
This follows directly from the fact that $S^*$ is accepted with
probability at least $1-\alpha$ since $H_{0,S^*}$ is true; see
\cite{PetBuhMei15} for details.

In the case of linear models, the method proposed by \citet[][Eq.~(16)]{PetBuhMei15} 
considers a set $S$ as invariant if there exist linear regression coefficients $\beta$ and error variance $\sigma$ which are identical across all environments. We consider the conditional independence relation in \eqref{eq:H0S} as a generalization, 
even for linear relations. In the following example the regression coefficients are the same in all environments, and the residuals have the same mean and variance, but differ in higher order moments \citep[cf.][Eq.~(3)]{PetBuhMei15}:

\begin{example} Consider a discrete environmental variable $E$. If in $E=1$ we have
$$
Y = 2X + N, N \independent X,
$$
and in $E=2$
$$
Y = 2X + M, M \independent X,
$$
where $M$ and $N$ have the same mean and variance but differ in higher order moments.
In this case, we would have
$E \notindependent Y \,|\, X$, but the hypothesis ``same linear regression coefficients and error variance'' cannot be rejected.
\end{example}

The question remains how to test~\eqref{eq:H0S}.
If we assume a linear function $f$ in the structural
equation~\eqref{eq:Y}, then tests that can guarantee the level as
in~\eqref{eq:level} are available \citep{PetBuhMei15}. 
The following examples show what could go wrong if the data 
contain nonlinearities that are 
not properly taken into account.
\begin{example}[Linear model and nonlinear data] \label{ex:counterlinear}
Consider the following SCM, in which $X_2$ and $X_3$ are direct causes of $Y$. 
\begin{align*}
X_1 &\leftarrow \envvar + \eta_X\\ 
X_2 &\leftarrow \sqrt{3X_1 + \eta_{X_1}}\\ 
X_3 &\leftarrow \sqrt{2X_1 + \eta_{X_2}}\\ 
Y &\leftarrow X_2^2 - X_3^2 + \eta_Y
\end{align*}
Due to the nonlinear effect, a linear regression from $Y$ on $X_2$ and $X_3$ does not yield an invariant model. 
If we regress $Y$ on $X_1$, however, we obtain invariant prediction and independent residuals. 
In this sense, the linear version of ICP fails but it still chooses a set of ancestors of $Y$ (it can be argued that this failure is not too severe).
\end{example}
\begin{example}[Linear model and nonlinear data] \label{ex:counterlinear2}
In this example, the model misspecification leads to a wrong set that includes a descendant of $Y$.
Consider the following SCM
\begin{align*}
X_1 &\leftarrow \envvar + \eta_1\\ 
Y &\leftarrow f(X_1) + \eta_Y\\ 
X_2 &\leftarrow g(X_1) + \gamma Y + \eta_{2}
\end{align*}
with independent Gaussian error terms.
Furthermore, assume that
\begin{align*}
\forall x \in \mathbb{R}: \;  \qquad f(x) &= \alpha x + \beta h(x) \\
\forall x \in \mathbb{R}: \; \qquad  g(x) &= h(x) - \gamma f(x)\\
 \beta \gamma^2  - \gamma &= - \beta \mathrm{var}(\eta_2) / \mathrm{var}(\eta_Y) 
\end{align*}
for some $\alpha, \beta$ and $h:\mathbb{R} \rightarrow \mathbb{R}$.
Then, in the limit of an infinite sample size, the set
$\{X_1, X_2\}$ is the only set that, after a linear regression, yields residuals that are independent of $\envvar$. 
(To see this write $Y = f(X_1) + \eta_Y$ as a linear function in $X_1$, $X_2$ and show that the covariance between the residuals and $X_2$ is zero.)
Here, the functions have to be ``fine-tuned'' in order to make the conditional $Y|X_1,X_2$ linear in $X_1$ and $X_2$.\footnote{This example is motivated by theory 
that combines linear and nonlinear models with additive noise
\citep{Ernest16}.}
As an example, one may choose $Y \leftarrow 
X_1 + 0.5X_1^2 + \eta_Y$ and 
$X_2 \leftarrow 0.5 X_1^2 - X_1 + Y + \eta_2$ and $\eta_1, \eta_Y, \eta_2$ i.i.d.\ with distribution $\mathcal{N}(0,\sigma^2 = 0.5)$.
\end{example}
The examples show that ICP loses its coverage guarantee if we assume linear relationships for testing~\eqref{eq:H0S} while the true data generating process is nonlinear.

In the general nonlinear and nonparametric case, however, it becomes more difficult to guarantee the type I error rate when testing the conditional independence~\eqref{eq:H0S} \citep{Shah2018}. This in contrast to nonparametric tests for (unconditional) independence \citep{bergsma2014consistent, szekely2007}.
In a nonlinear conditional independence test setting, where we know an
appropriate parametric basis expansion for the causal effect of the variables we
condition on, we can of course revert back to unconditional
independence testing.
Apart from such special circumstances, we have to find tests that guarantee the type I error rate in~\eqref{eq:level} as closely as possible under a wide range of scenarios. 
We describe some methods that test~\eqref{eq:H0S} in Section~\ref{subsec:condtest} but for now let us assume that we are given such a test. 
We can then apply the method of nonlinear ICP~\eqref{eq:Shat} to the example of fertility data.

\paragraph{Example (Fertility data).} The following sets were accepted at the level $\alpha = 0.1$ 
when using nonlinear ICP with invariant conditional quantile prediction 
(see Appendix~\ref{supp:sec:condtest_details} for details) as a conditional independence test: 
{\small
\begin{align*}
S_1 &= \lbrace \text{Q5} \rbrace \\
S_2 &= \lbrace \text{IMR, Imports of goods and services, Urban pop.\ (\% of total)} \rbrace \\
S_3 &= \lbrace \text{IMR, Education expend. (\% of GNI), Exports of goods and services, GDP per capita} \rbrace
\end{align*}
}
As the intersection of $S_1, \ldots, S_{3}$ is empty, we have $\hat{S} = \emptyset$. This motivates the concept of defining sets.

\subsection{Defining sets} \label{subsec:defining}
It is often impossible to distinguish between highly correlated
variables. For example, infant mortality \text{IMR} and under-five mortality \text{Q5} are
highly correlated 
in the data and can often be substituted for each
other. We accept sets that contain either of these variables. When taking the intersection as in~\eqref{eq:Shat}, this leads to exclusion of both variables in $\hat{S}$ and
potentially to an altogether empty set $\hat{S}$. We can instead ask
for the defining sets \citep{goeman2011multiple}, where  a defining set $\hat{D}\subseteq\{1,\ldots,p\}$ has the
properties
\begin{enumerate}[(i)]
\item $S \cap \hat{D} \neq \emptyset$ for all $S$ such that $H_{0,S}$ is
  accepted.
\item there exists no strictly smaller set $D'$ with $D' \subset 
\hat{D}$ 
for which property
  (i) is true.
\end{enumerate}
In words, we are looking for subsets $\hat{D}$, such that each accepted set $S$ has at least one element that also appears in $\hat{D}$.
If the intersection $\hat{S}$~\eqref{eq:Shat} is non-empty, any subset
of $\hat{S}$ that contains only one variable is a defining
set. Defining set are especially useful, however, in cases where the
intersection $\hat{S}$ \emph{is} empty. We still know that, with high probability, at least one
of the variables in the defining set $\hat{D}$ has to be a parent.
Defining sets are not necessarily unique.
Given a defining set $\hat{D}$, we thus know that
 \[ P( S^* \cap \hat{D}
=\emptyset) \leq 
P(H_{0,S^*} \text{ rejected}) 
\leq \alpha.\] 
That is,
a) at least one of the variables in the defining
set $\hat{D}$ is a parent of the target, and b) the data do not
allow to resolve it on a finer scale. 

\paragraph{Example (Fertility data).} We obtain seven defining sets:
{\small
\begin{align*}
\hat{D}_1 &= \lbrace \text{IMR, Q5} \rbrace \\
\hat{D}_2 &= \lbrace \text{Q5, Education expenditure (\% of GNI), Imports of goods and services} \rbrace \\
\hat{D}_3 &= \lbrace \text{Q5, Education expenditure (\% of GNI), Urban pop.\ (\% of total)} \rbrace \\
\hat{D}_4 &= \lbrace \text{Q5, Exports of goods and services, Imports of goods and services} \rbrace \\
\hat{D}_5 &= \lbrace \text{Q5, Exports of goods and services, Urban pop.\ (\% of total)} \rbrace \\
\hat{D}_6 &= \lbrace \text{Q5, GDP per capita, Imports of goods and services} \rbrace \\
\hat{D}_7 &= \lbrace \text{Q5, GDP per capita, Urban pop.\ (\% of total)} \rbrace
\end{align*}}
Thus the highly-correlated variables infant mortality \text{IMR} and
under-five mortality \text{Q5} indeed form one of the defining sets in
this example in the sense that we know at least one of the two is a
causal parent for fertility but we cannot resolve which one it is or
whether both of them are parents. 

\subsection{Confidence bands}\label{subsec:bands}
For a given  set $S$, we can in general construct a $(1-\alpha)$-confidence band $\hat{\mathcal{F}}_S $ for the regression function when predicting $Y$ with the variables $X_S$. 
Note that if
$f$ is the regression function 
when regressing $Y$ on the true set of causal variables $X_{S^*}$ 
and hence,
then, with probability $1-\alpha$, we have 
\[ P( f\in \hat{\mathcal{F}}_{S^*}) \ge 1-\alpha .\]
Furthermore, from Section~\ref{subsec:nonlinICP} we know that $H_{0,S^*}$ is accepted with probability $1-\alpha$. We can hence construct a confidence band for the causal effects as 
\begin{equation}\label{eq:hatF}  \hat{\mathcal{F}} := \bigcup_{S:H_{0,S} \text{ not rejected}} \hat{\mathcal{F}}_S .\end{equation}
Using a Bonferroni correction, we have the guarantee that 
 \[ P\big( f \in \hat{\mathcal{F}}\big) \; \ge\; 1-2\alpha ,\]
where
the coverage guarantee is point-wise or uniform, depending on the
coverage guarantee of the underlying estimators $\hat{\mathcal{F}}_{S}$ for all
given $S\subseteq\{1,\ldots,p\}$.

\subsection{Average causal effects}\label{subsec:prediction}

The confidence bands $\hat{\mathcal{F}}$ themselves can be difficult
to interpret. 
Interpretability can be guided by looking at the average causal effect
in the sense that we compare  the
expected response at  $\tilde{x}$ and $x $:
\begin{equation} \label{eq:Delta} 
\ACE(\tilde{x},x) \; := \; E\big( Y\big| \mbox{do}(X=\tilde{x}) \big) -E\big( Y\big| \mbox{do}(X={x}) \big).
\end{equation}
For the fertility data, this would involve a hypothetical scenario
where we fix the variables to be equal to ${x}$ for a country in the second term and, for the first term,
we set the variables to $\tilde{x}$, which might differ from $x$ just in one or a
few coordinates. Eq.~\eqref{eq:Delta} then compares the average expected fertility between
these two scenarios. Note that the expected response under a
do-operation is just a function of the causal variables 
$S^* \subseteq \{1,\ldots,p\}$. That is---in the absence of hidden
variables---we have 
\[ E\big( Y\big| \mbox{do}(X=x) \big) \;=\; E\big( Y\big|
\mbox{do}(X_{S^*}=x_{S^*}) \big) ,\] and  the latter is then equal to 
\[ E\big( Y\big|
\mbox{do}(X_{S^*}=x_{S^*}) \big) \; =\;  E\big( Y\big|
X_{S^*}=x_{S^*} \big) ,\] that is it does not matter whether we set the
causal variables to a specific value $x_{S^*}$ or whether they were observed in
this state.

Once we have a confidence band as defined in~\eqref{eq:hatF},
we can bound the average causal effect~\eqref{eq:Delta}  by the interval 
\[ \widehat{\ACE}(\tilde{x},x) := \big[  \inf_{g\in \hat{\mathcal{F}}} ( g(\tilde{x}) -g(x) ) , \; \sup_{g\in \hat{\mathcal{F}}} ( g(\tilde{x}) -g(x) )  \big] ,\]
with the immediate guarantee that 
\begin{equation}\label{eq:2a} P\big(   \ACE(\tilde{x},x) \;\in \; \widehat{\ACE}(\tilde{x},x) \big) \ge 1-2\alpha, \end{equation}
where the factor $2\alpha$ is guarding, by a Bonferroni correction, against both a probability $\alpha$ that $S^*$ will not
be accepted---and hence $\hat{S}\subseteq S^*$ is not necessarily true---and another probability $\alpha$ that the confidence bands
will not  provide coverage for the  parental set $S^*$.

\paragraph{Example (Fertility data).}  
The confidence bands $\hat{\mathcal{F}}$, required for the computation
of $\widehat{\ACE}(\tilde{x},x)$, are obtained by a time series bootstrap
\citep{kunsch1989jackknife} as the fertility data contain temporal dependencies. The time series bootstrap procedure is described in Appendix~\ref{supp:tsboots}. We use a level of $\alpha = 0.1$ which
implies a coverage guarantee of 80\% as per \eqref{eq:2a}. In the examples below, we set $x$ to an observed data point and vary only $\tilde{x}$.

\begin{figure*}[!t]
\subfloat[{\footnotesize $\widehat{\ACE}_{\text{IMR}}(\tilde{x},x)$} ]{
    \includegraphics[page = 1, width=0.3\textwidth, keepaspectratio=true]{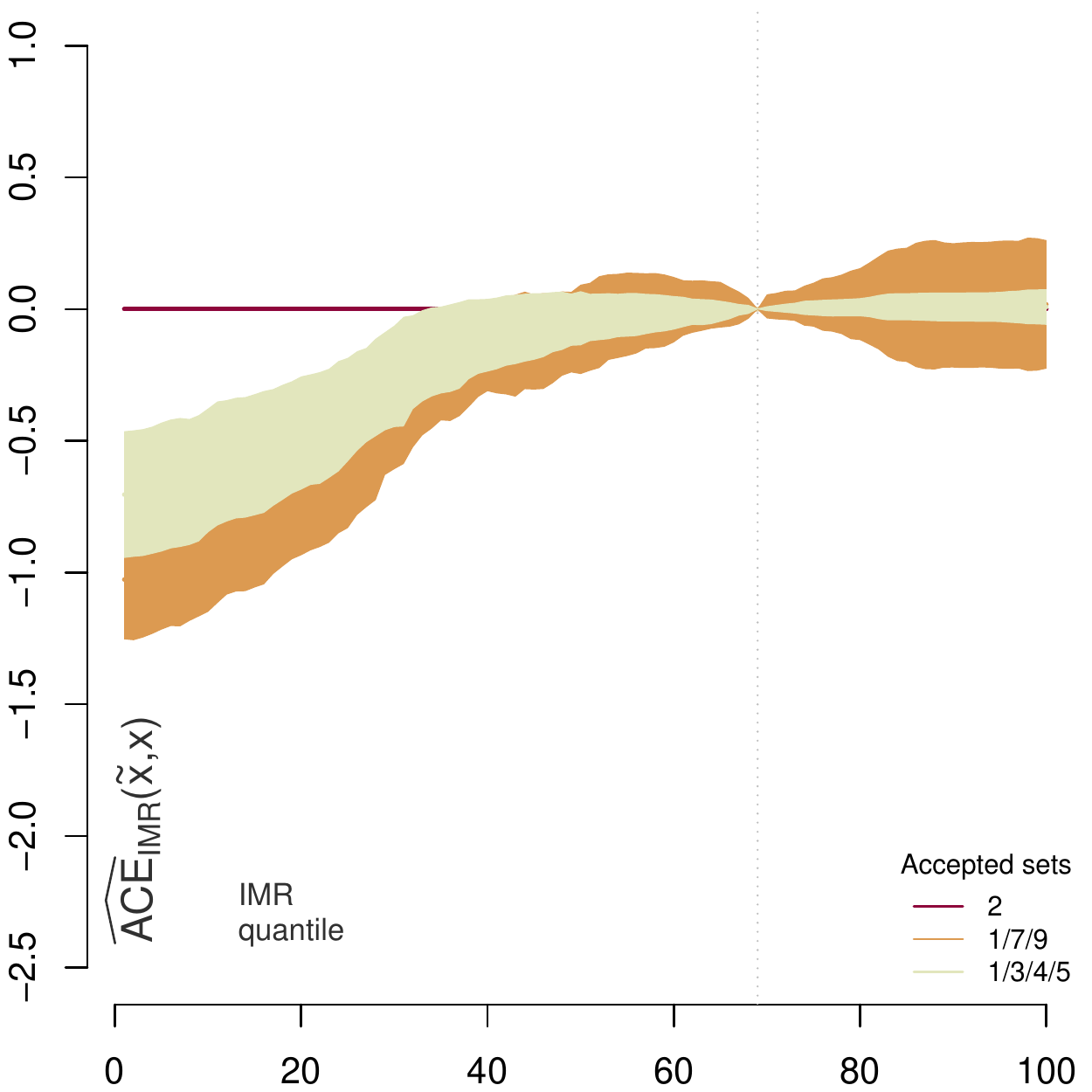}
    \label{fig:def_sets_ind1}
}
\hspace{.25cm}
\subfloat[{\footnotesize $\widehat{\ACE}_{\text{Q5}}(\tilde{x},x)$}]{
    \includegraphics[page = 2, width=0.3\textwidth, keepaspectratio=true]{figures/quantRfExceedance_fishersTestExceedance_methodrandomForest_jointlyFALSE_baseline_setToPoint_pointToSetTo728Nigeria_1993_Africa_confbootsConf_logTRUE_nsim50_degreeFit_nSubFrac_varsDirMapUsedFALSE}
    \label{fig:def_sets_ind2}
}
\hspace{.25cm}
\subfloat[{\footnotesize $\widehat{\ACE}_{\text{IMR}+\text{Q5}}(\tilde{x},x)$}]{
    \includegraphics[width=0.3\textwidth, keepaspectratio=true]{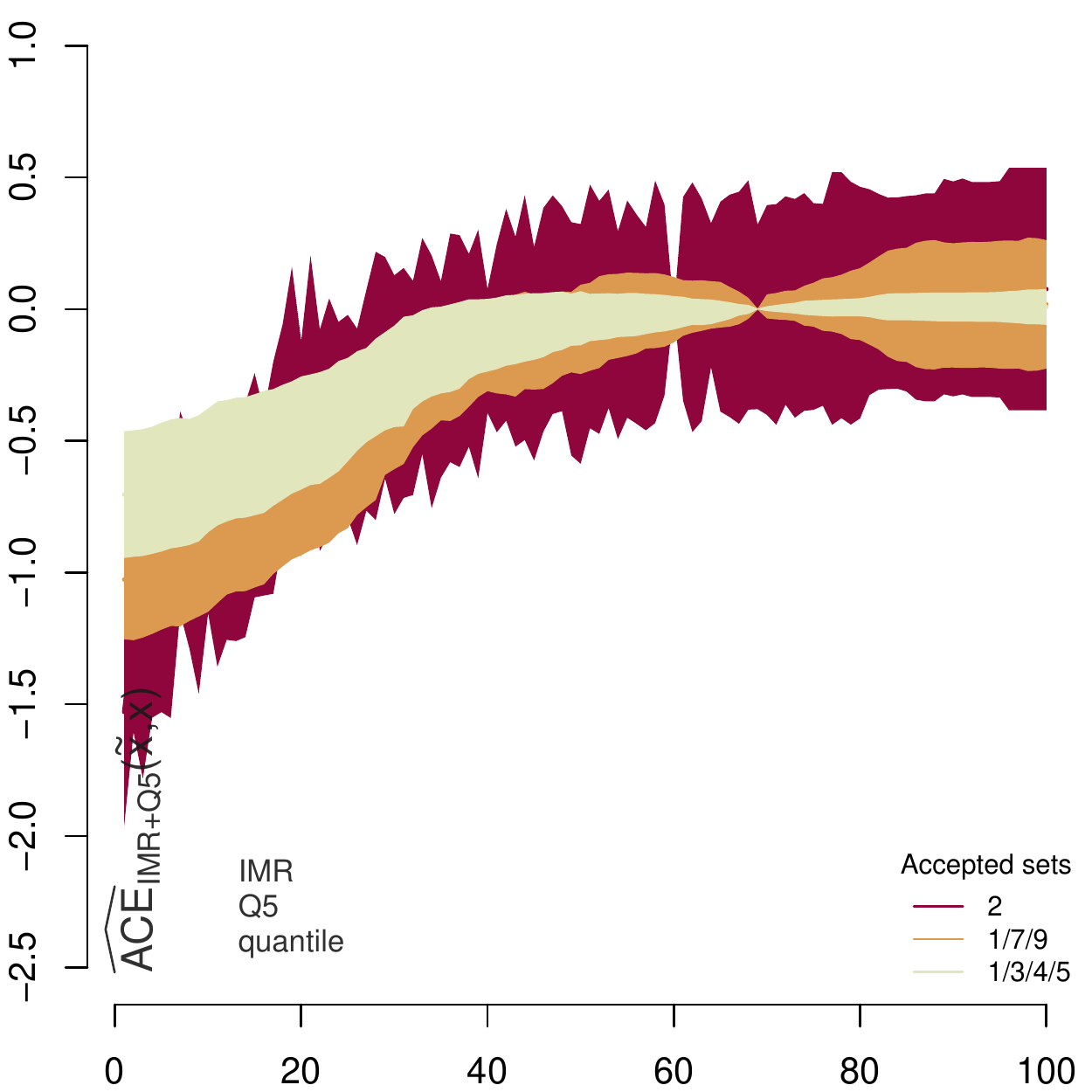}
    \label{fig:def_sets_jointly}
}
\vspace{-.15cm}
\caption{\small  Data for Nigeria in 1993: The union of the 
confidence
  bands $\hat{\mathcal{F}}_S$, denoted by $\hat{\mathcal{F}}$, bounds the average causal effect of varying the variables in the defining set $\hat{D}_1 = \lbrace \text{IMR, Q5} \rbrace$ on the target $\log(\text{TFR})$. IMR and~Q5 have been varied individually, see panels \protect\subref{fig:def_sets_ind1} and \protect\subref{fig:def_sets_ind2}, as well as jointly, see panel~\protect\subref{fig:def_sets_jointly}, over their respective quantiles. 
  In panels \protect\subref{fig:def_sets_ind1} and \protect\subref{fig:def_sets_ind2}, we do not observe consistent effects
different from zero as some of the accepted models do not contain IMR and some do not contain Q5. However, when varying the variables $\hat{D}_1 =\lbrace \text{IMR, Q5} \rbrace$ jointly (see panel \protect\subref{fig:def_sets_jointly}), we see that all accepted models predict an increase in expected $\log(\text{TFR})$ as \text{IMR} and \text{Q5} increase.  
\label{fig:intervention_effects} }
\end{figure*}

In the first example, we consider the observed covariates for Nigeria
in 1993 as ${x}$.
The point of comparison $\tilde{x}$ is set equal to
${x}$, except for the variables in the defining set $\hat{D}_1 = \lbrace \text{IMR, Q5} \rbrace$. 
In Figures~\ref{fig:intervention_effects}\subref{fig:def_sets_ind1} 
and~\subref{fig:def_sets_ind2}, these are varied 
individually over their respective quantiles. 
The overall confidence interval $\hat{\mathcal{F}}$ consists of the union of the shown confidence intervals $\hat{\mathcal{F}}_S$.
If $x=\tilde{x}$ (shown by the vertical lines), the
average causal effect is zero, of course. 
In neither of the two
scenarios
shown in Figures~\ref{fig:intervention_effects}\subref{fig:def_sets_ind1} and~\subref{fig:def_sets_ind2}, we observe consistent effects
different from zero as some of the accepted models do not contain IMR
and some do not contain Q5. 
However, when varying the variables $\hat{D}_1 =
\lbrace \text{IMR, Q5} \rbrace$
jointly (see Figure~\ref{fig:intervention_effects}\subref{fig:def_sets_jointly}),
we see that all accepted models predict an increase in
expected 
$\log(\text{TFR})$ as \text{IMR} and \text{Q5} increase.

\begin{figure*}[!t]
\center
\subfloat[\footnotesize $\widehat{\ACE}_{\text{IMR}+\text{Q5}}(\tilde{x},x)$
estimated using nonlinear ICP with invariant conditional quantile
prediction]{
     \includegraphics[page = 1, width=1\textwidth, keepaspectratio=true]{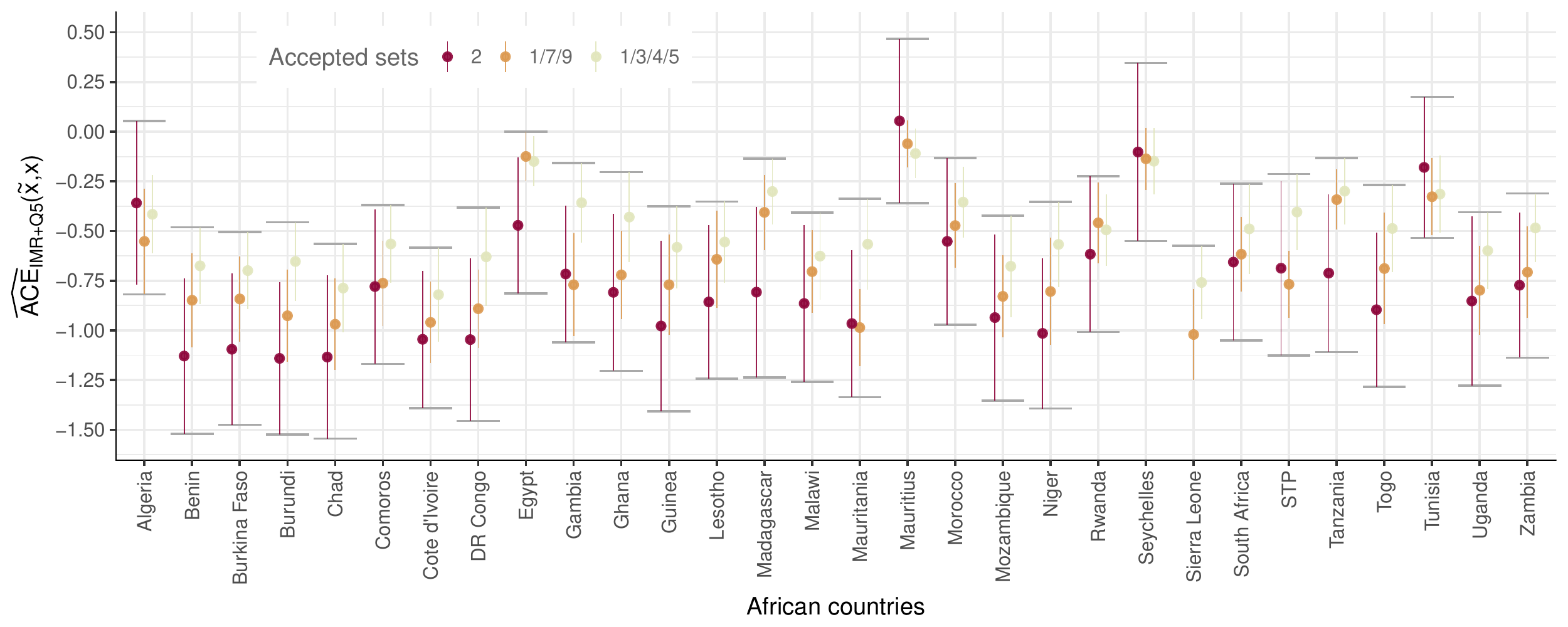}
         \label{fig:ace}
}

\subfloat[\footnotesize Analogous computation as in \protect\subref{fig:ace} \textit{assuming} that \textit{all} covariates have a direct causal effect on the target, using a Random Forest regression model with bootstrap confidence intervals]{
    \includegraphics[page = 1, width=1\textwidth, keepaspectratio=true]{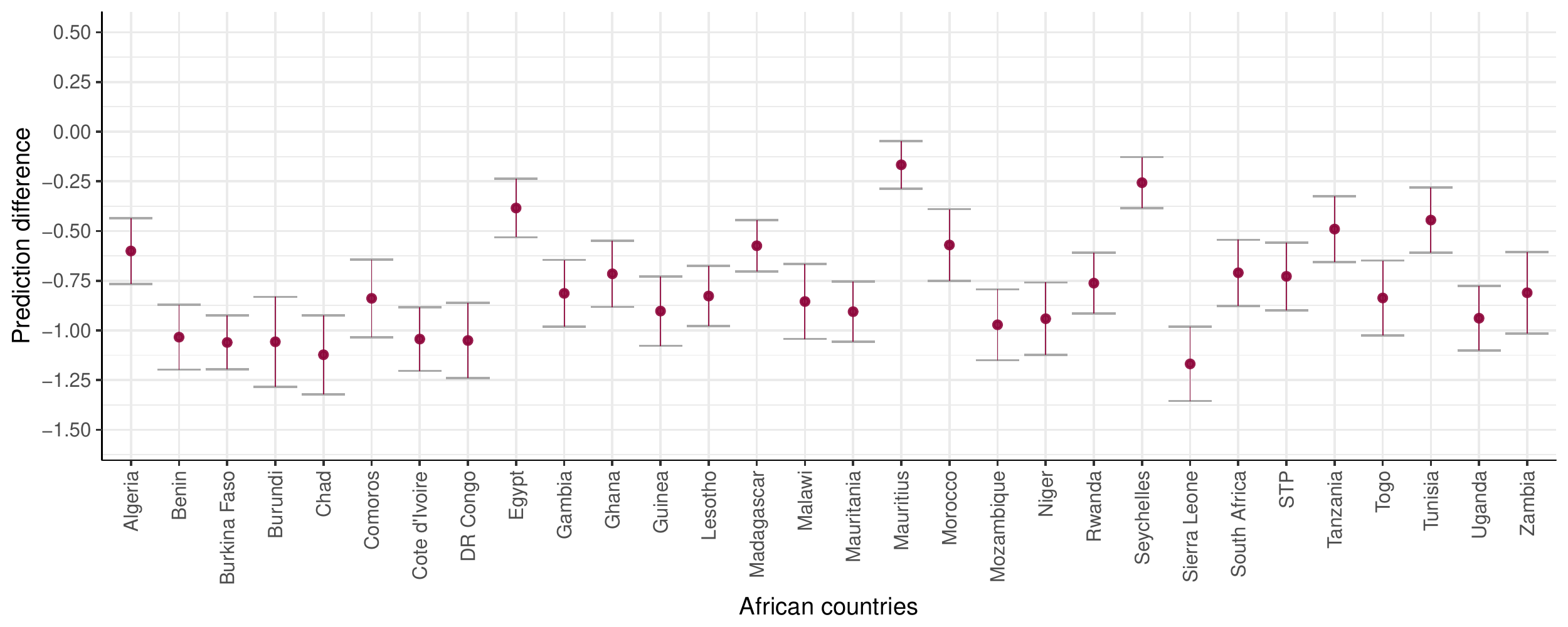}
    \label{fig:rf}
}
\caption{\small \protect\subref{fig:ace} Bounds for the average causal
  effect of setting the variables IMR and Q5 in the African countries
  in 2013 to European levels, that is $\tilde{x}$ differs from the
  country-specific observed values $x$ in
that the child mortality rates $\text{IMR}$ and $\text{Q5}$ have been
set to their respective European average. The implied coverage guarantee is
80\% as we chose $\alpha=0.1$.  \protect\subref{fig:rf}
Random Forest regression model using all covariates as input. The
(non-causal)  regression effect coverage is again set to 80\%.
We will argue below that the confidence intervals obtained by random forest are too small, see Table~\ref{tab:datasets} and Figure~\ref{fig:cv_cov_all}.
\label{fig:intervention_effects2} }
\end{figure*}

In the second example, we compare the expected fertility rate between
countries where all covariates are set to the value ${x}$, which is
here chosen to be equal to the observed values of all African
countries in 2013. The expected value of log-fertility under this value ${x}$ of
covariates is compared to the scenario where we take as $\tilde{x}$
the same value but set the values of 
 the child-mortality variables IMR and Q5 to their respective
European averages. The union of intervals in
Figure~\ref{fig:intervention_effects2}\subref{fig:ace} (depicted by the horizontal line segments) 
correspond to  $\widehat{\ACE}(\tilde{x},x)$ for each
country under 
nonlinear ICP with invariant conditional quantile prediction. The
accepted models make largely coherent predictions for the effect associated with
this comparison. For most
countries, the difference is negative, meaning that 
the average expected fertility declines if the child mortality rate in a country 
decreases to European levels. The
countries where $\widehat{\ACE}_{\text{IMR}+\text{Q5}} (\tilde{x},x)$ contains~0 typically have
a child mortality rate that is close to European levels, meaning that there
is no substantial difference between the two points $\tilde{x},x$ of comparison.

For comparison, in Figure~\ref{fig:intervention_effects2}\subref{fig:rf}, we show the equivalent computation as in Figure~\ref{fig:intervention_effects2}\subref{fig:ace} when {\it all} covariates are {\it assumed} to have a direct causal effect 
on the target and a Random Forest is used for estimation \citep{breiman01random}.  We observe that while the resulting regression bootstrap confidence intervals often overlap with $\widehat{\ACE}_{\text{IMR}+\text{Q5}}(\tilde{x},x)$, they are typically much smaller. This implies that if the regression model containing all covariates was---wrongly---used as a surrogate for the causal model, 
the uncertainty of the prediction would be underestimated. Furthermore, 
such an approach ignoring the causal structure 
can lead to a significant bias in the prediction of causal effects
when we consider interventions on descendants of the target variable, for example.

\begin{table}
    \caption{Coverage\label{tab:coverage}}
    \label{tab:datasets}
\begin{center}
\begin{small}
\begin{tabular}{lccccc}
\hline
\abovespace\belowspace
Coverage guarantee & 0.95 & 0.90 & 0.8 & 0.5\\
\hline
\abovespace
Coverage with nonlinear ICP & 0.99 & 0.95 & 0.88 & 0.58 \\
\belowspace
Coverage with Random Forest & 0.76 & 0.71 & 0.61 & 0.32 \\
\belowspace
Coverage with mean change & 0.95 & 0.88 & 0.68 & 0.36 \\
\hline
\end{tabular}
\end{small}
\end{center}
\end{table}

\begin{figure*}[!t]
\center
    \includegraphics[page = 1, width=1\textwidth, keepaspectratio=true, trim = {0 0 0 25}, clip]{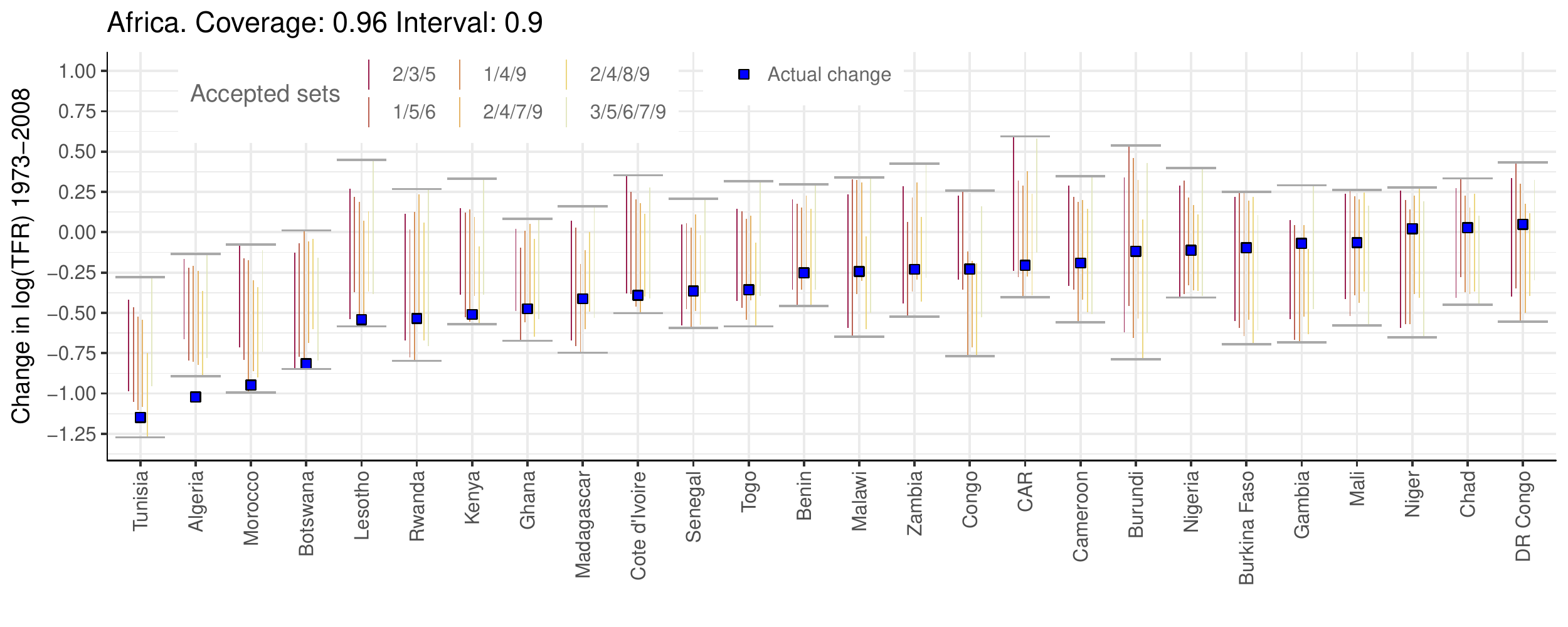}
\caption{\small\label{fig:cv_cov_africa} The confidence intervals show the predicted change in $\log{(\text{TFR})}$ from 1973 -- 2008 for all African countries when not using their data in the nonlinear ICP estimation (using invariant conditional quantile prediction with $\alpha = 0.1$). In other words, only the data from the remaining five continents was used during training. The horizontal line segments mark the union over the accepted models' intervals for the predicted change; the blue squares show the true change in $\log{(\text{TFR})}$ from 1973 -- 2008. Only those countries are displayed for which the response was not missing in the data, i.e.\ where  $\log{(\text{TFR})}$ in 1973 and in 2008 were recorded. The coverage is $25/26 \approx 0.96$.}
\end{figure*}

Lastly, we consider a cross validation scheme over time to assess the coverage properties of nonlinear ICP. We leave out
the data corresponding to one entire continent and run nonlinear ICP
with invariant conditional quantile prediction using the data from the
remaining five continents. 
\begin{figure*}[!t]
\center
\subfloat[\footnotesize Nonlinear ICP with invariant conditional quantile prediction. Coverage: 0.88.]{
     \includegraphics[page = 1, width=.975\textwidth, keepaspectratio=true, trim = {0 0 0 25}, clip]{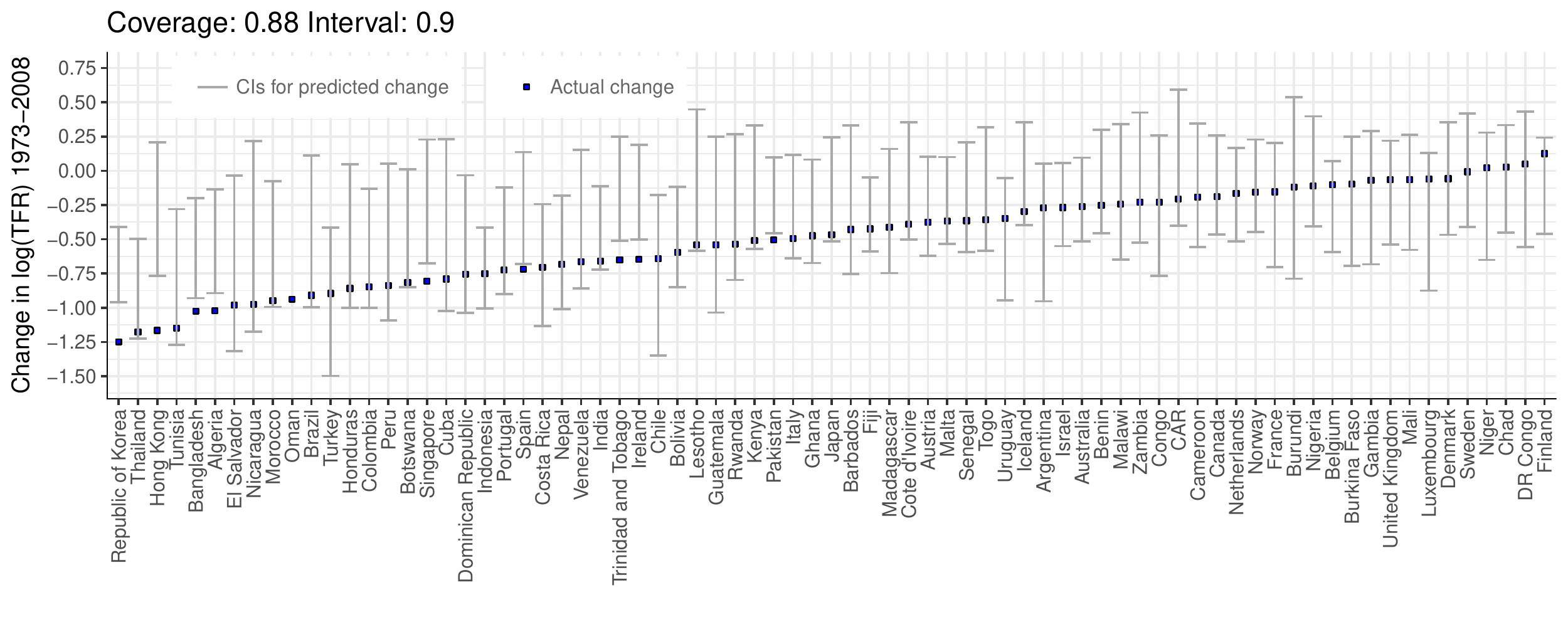}
         \label{fig:icp_cv_cov}
}

\subfloat[\footnotesize Random Forest regression model with bootstrap confidence intervals. Coverage: 0.61.]{
    \includegraphics[page = 1, width=.975\textwidth, keepaspectratio=true, trim = {0 0 0 25}, clip]{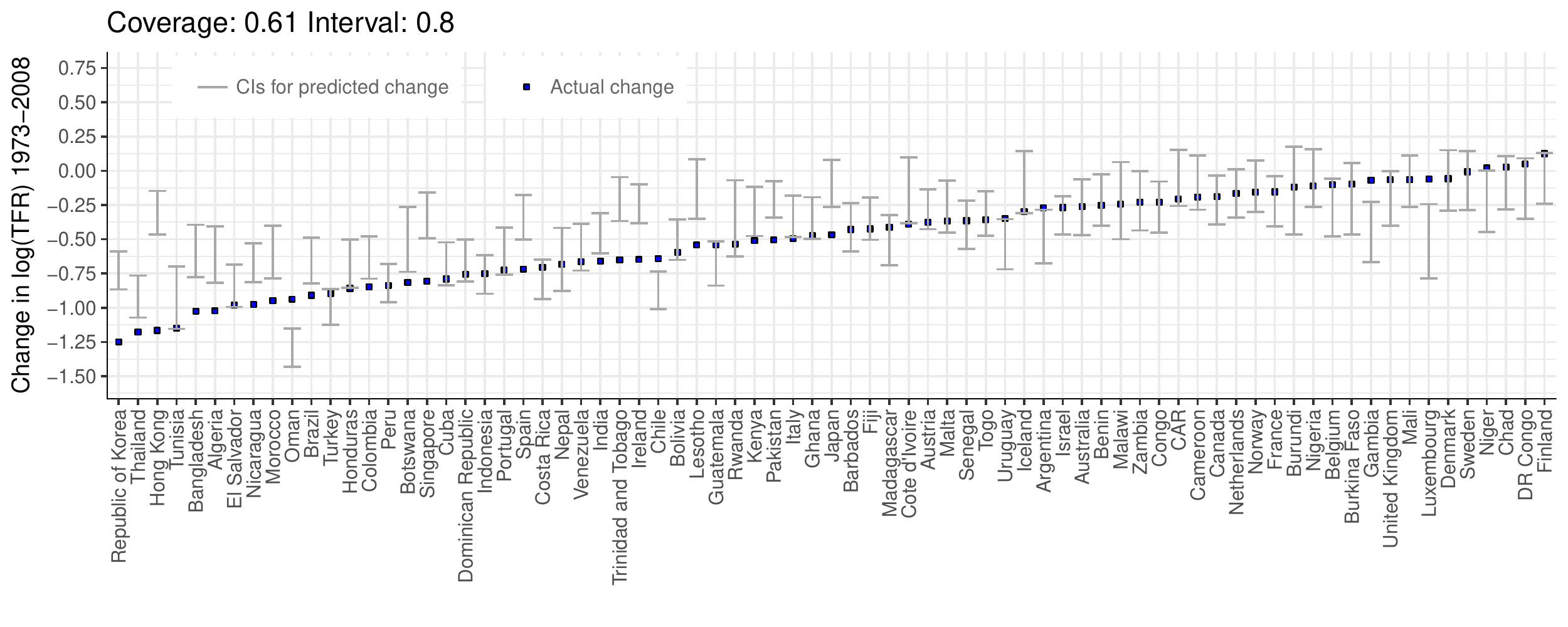}
    \label{fig:rf_cv_cov}
}

\subfloat[\footnotesize Prediction based on mean change on continents other than country's own continent. Coverage: 0.68.]{
    \includegraphics[page = 1, width=0.975\textwidth, keepaspectratio=true, trim = {0 0 0 25}, clip]{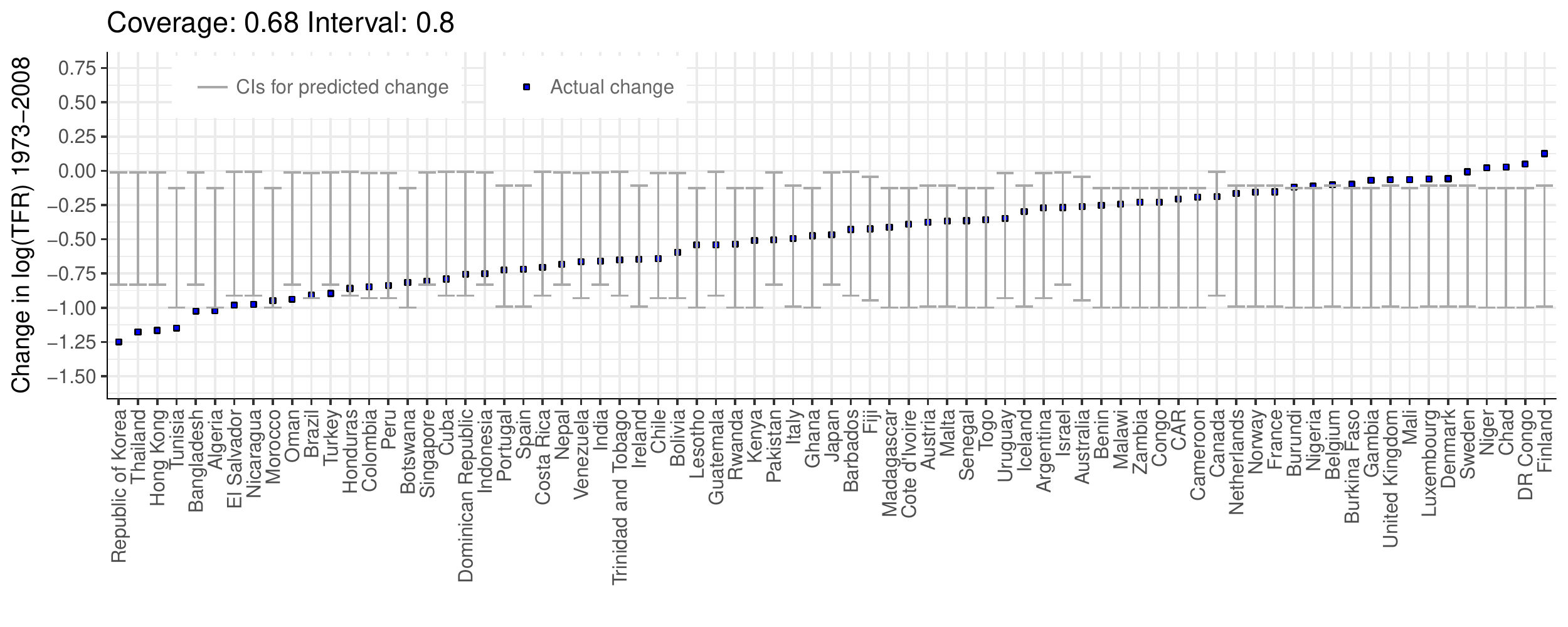}
    \label{fig:mean_change_cv_cov}
}

\caption{\small\label{fig:cv_cov_all} The confidence intervals show
  the predicted change in $\log{(\text{TFR})}$ from 1973 -- 2008 for
  all countries when not using the data of the country's continent in
  the estimation (with implied coverage
  guarantee 80\%). Only those countries are displayed for which $\log{(\text{TFR})}$ in 1973 and 2008 were not missing in the data. For nonlinear ICP the shown intervals are the union over the accepted models' intervals.}
\end{figure*}
We perform this leave-one-continent-out
scheme for different values of $\alpha$. For each value of $\alpha$,
we then compute the predicted change in the response
$\log{(\text{TFR})}$ from 1973 -- 2008 for each country belonging to
the continent that was left out during the estimation procedure. The
predictions are obtained by using the respective accepted
models.\footnote{Their number differs according to $\alpha$: for a
  smaller $\alpha$, additional models can be accepted compared to
  using a larger value of $\alpha$. In other words, the accepted
  models for $\alpha_2$ where $\alpha_1 < \alpha_2$ are a subset of
  the accepted models for $\alpha_1$.} We then compare the union of
the associated confidence intervals with the real, observed change in
$\log{(\text{TFR})}$. This allows us to compute the coverage
statistics shown in Table~\ref{tab:coverage}. 
We observe that
nonlinear ICP typically achieves more accurate coverage compared to (i) a Random Forest regression model including all variables and (ii) a baseline where the predicted change in $\log{(\text{TFR})}$ for a country is the observed mean change in $\log{(\text{TFR})}$ across all continents other than the country's own continent. Figures~\ref{fig:cv_cov_africa} and~\ref{fig:cv_cov_all} show the confidence intervals and the observed values for all African countries (Figure~\ref{fig:cv_cov_africa}) and all countries (Figure~\ref{fig:cv_cov_all}) with observed $\log{(\text{TFR})}$ in 1973 and 2008. 

Recall that one advantage of a causal model is that, in the absence of hidden variables, it does not matter whether certain variables have been intervened on or whether they were observed in this state -- the resulting prediction remains correct in any of these cases. On the contrary, the predictions of a non-causal model can become drastically incorrect under interventions. This may be one reason for the less accurate coverage statistics of
the Random Forest regression model---in this example, it seems
plausible that some of the predictors were subject to different
external `interventions' across continents and countries.

\section{Conditional Independence Tests}\label{subsec:condtest}
We present and evaluate an array of methods for testing conditional independence in a nonlinear setting, many of which exploit the invariance of causal models across different environments. Here, we briefly sketch the main ideas of the considered tests, their respective assumptions and further details are provided in Appendix~\ref{supp:sec:condtest_details}. All methods (A) -- (F) are available in the package \texttt{CondIndTests} for the \textsf{R} language.
Table~\ref{tab:package} in Appendix~\ref{supp:ssec:package} shows the supported methods and options. 
An experimental comparison of the corresponding power and type I error rates of these tests can be found in Section~\ref{sec:experiments}.

\begin{enumerate}[(A)]
\item 
{\bf Kernel conditional independence test.} 
Use a kernel conditional independence test for $Y \independent \envvar \;
  \vert \; \Xs$ \citep{Fukumizu2008, Zhang11}. See Appendix~\ref{supp:ssec:kernel} for further details.

\item 
{\bf Residual prediction test.} 
Perform a nonlinear regression from $Y$ on $X_S$, using an appropriate basis expansion, and apply a variant of a Residual Prediction (RP) test \citep{Shah15}. The main idea is to scale the residuals of the regression such that the resulting test statistic is not a function of the unknown noise variance. 
This allows for a straight-forward test for dependence between the residuals and ($E$, $X_S$). In cases where a suitable basis expansion is unknown, random features \citep{Williams2000, Rahimi2007} can be used as an approximation. See Appendix~\ref{supp:ssec:rptest} for further details.

\item 
{\bf Invariant environment prediction.} 
Predict the environment $E$, once with a model that uses $X_S$
  as predictors only and once
  with a model that uses $(X_S,Y)$ as predictors. If the null is true and we find the optimal model in
  both cases, then the out-of-sample performance of both models is
  statistically indistinguishable. See Appendix~\ref{supp:ssec:envpred} for further details.

\item 
{\bf Invariant target prediction.} 
Predict the target $Y$, once with a model that uses  $X_S$
  as predictors only and once
  with a model that uses $(X_S,E)$ as predictors. If the null is true and we find the optimal model in
  both cases, then the out-of-sample performance of both models is
  statistically indistinguishable. See Appendix~\ref{supp:ssec:targetpred} for further details.

\item 
{\bf Invariant residual distribution test.} 
Pool the data across all environments and predict the response $Y$ with variables $X_S$.
 Then test whether the
  distribution of the residuals is identical in all environments $E$.  
See Appendix~\ref{supp:ssec:residdis} for further details.  
\item 
{\bf Invariant conditional quantile prediction.} Predict a $1-\beta$ quantile of the conditional distribution of $Y$, given $X_S$, by pooling the data over all environments. Then test whether the exceedance of the conditional quantiles is independent of the environment variable. Repeat for a number of quantiles and aggregate the resulting individual $p$-values by Bonferroni correction.
See Appendix~\ref{supp:ssec:invquant} for further details.
\end{enumerate}

Another interesting possibility for future work would be to devise a conditional independence test based on model-based recursive partitioning \citep{Zeileis2008, R-partykit}.

Non-trivial, assumption-free conditional independence tests with a valid
level do not exist \citep{Shah2018}.
It 
is therefore not surprising that 
all of the above tests 
assume 
the dependence on the conditioning variable to be ``simple'' in one form or the other. 
Some of the above tests require the noise variable in~\eqref{eq:Y} to be additive in the sense that we do not expect the respective test to have the correct level when the noise is not additive. As additive noise is also used in Sections~\ref{subsec:bands} and \ref{subsec:prediction}, we have written the structural equations above in an additive form.

One of the inherent difficulties with these tests is that the
estimation bias when conditioning on potential parents in~\eqref{eq:H0S} 
can potentially lead to a more frequent rejection of a true null hypothesis than the nominal level suggests. In approaches (C) and (D), we also need to test whether the predictive accuracy is identical under both models and in approaches (E) and (F) we need to test whether univariate distributions remain invariant across environments. While these additional tests are relatively straightforward, a choice has to be made.  

\paragraph{Discussion of power.}
Conditional independence testing is a statistically challenging problem. For the setting where we condition on a continuous random variable, we are not aware of any conditional independence test that holds the correct level and still has (asymptotic) power against a wide range of alternatives. 
Here, we want to briefly mention some power properties of the tests we have discussed above.

Invariant target prediction (D), for example, has no power to detect
if the noise variance is a function of $E$, as shown by the following example
\begin{example}
Assume that the distribution is entailed by the following model
\begin{align*}
E &\leftarrow  0.2\eta_E\\
X &\leftarrow  \eta_X\\
Y &\leftarrow  X^2 + E\cdot \eta_Y,
\end{align*}
where $\eta_E, \eta_X, \eta_Y \overset{\text{i.i.d.}}{\sim} \mathcal{N}(0,1)$. Then, any regression 
from 
$Y$ on $X$ and $E$ yields the same results as regressing $Y$ on $X$ only. That is, 
$$
\text{for all } x, e:\; \mean[Y \given X = x] = \mean[Y \given X = x, E = e]
$$
although 
$$
Y \notindependent E \given X.
$$
\end{example}

The invariant residual distribution test (E), in contrast, assumes homoscedasticity and might
have wrong coverage if this assumption is violated. 
Furthermore, two different linear models do not necessarily yield different distributions of the residuals when performing a regression on the pooled data set. 
\begin{example} \label{ex:counterrr}
Consider the following data generating process
\begin{align*}
Y^{e=1} &\leftarrow  2X^{e=1} + N^{e=1}\\
Y^{e=2} &\leftarrow  -X^{e=2} + 0.3N^{e=2},
\end{align*}
where the input variables $X^{e=1}$ and $X^{e=2}$ and the noise variables $N^{e=1}$ and $N^{e=2}$ have the same distribution in each environment, respectively. Then, approach (E) will accept the null hypothesis of invariant prediction. 
\end{example}
It is possible to reject the null hypothesis of invariant prediction in Example~\ref{ex:counterrr} by testing whether in each environment the residuals are uncorrelated from the input.

Invariant conditional quantile prediction (F) assumes
neither homoscedasticity nor does it suffer from the same issue of
(D), i.e.\ no power against an alternative where the noise variance $\sigma$ 
is a function of $E$. However, it is possible to construct examples
where (F) will have no power if the noise variance is a function of
both $E$ \emph{and} the causal variables $X_{S^*}$. Even then, though,
the  noise level would have to be carefully balanced to reduce the
power to 0 with approach (F) as the exceedance probabilities of
various quantiles (a function of $X_{S^*}$) would have to remain constant if we condition on
various values of $E$.

\section{Simulation Study}\label{sec:experiments}

\begin{figure}[ht]
\begin{center}
\begin{tikzpicture}[scale=1, line width=0.5pt, minimum size=0.58cm, inner sep=0.3mm, shorten >=1pt, shorten <=1pt]
    \normalsize
    \draw (2,2) node(2) [minimum size=6mm,rounded corners=2mm, draw] {\small 2};
    \draw (0,0) node(3) [minimum size=6mm,rounded corners=2mm, draw]   {\small 3};
    \draw (-2,2) node(1) [minimum size=6mm,rounded corners=2mm, draw]   {\small 1};
    \draw (2,0) node(5) [minimum size=6mm,rounded corners=2mm, draw] {\small 5};
    \draw (-2,-2) node(4) [minimum size=6mm,rounded corners=2mm, draw] {\small 4};
    \draw (2,-2) node(6) [minimum size=6mm,rounded corners=2mm, draw] {\small 6};
    \draw [-arcsq] (1) -- node[above] {+}(2);
    \draw [-arcsq] (1) -- node[left] {+} (3);
    \draw [-arcsq] (2) -- node[above] {$-$} (3);
    \draw [-arcsq] (3) -- node[left] {$-$}  (4);
        \draw [-arcsq] (3) -- node[above] {+} (6);
    \draw [-arcsq] (4) -- node[above] {$-$}  (6);
    \draw [-arcsq] (5) -- node[right] {+}  (6);
   \end{tikzpicture}
\caption{\small  The structure of the causal graph used in the simulations. The causal order is unknown for the simulations. All edge weights are 1 in absolute value.
 }
 \label{fig:dag_sim}
\end{center}
\end{figure}
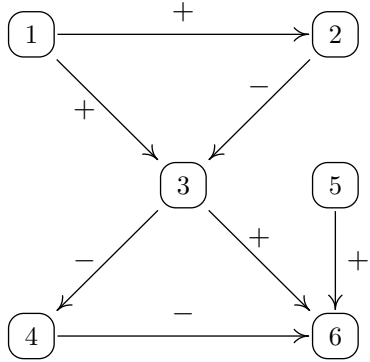

For the simulations, we generate data from different nonlinear additive noise causal models and compare the performance of the proposed conditional independence tests.  The structural equations are of the form 
$Z_k  \;\leftarrow \;  g_k(Z_{\text{pa}_k}) + \eta_k, $ where the structure of the DAG is shown in Figure~\ref{fig:dag_sim} and kept constant throughout the simulations for ease of comparison. 
We vary the nonlinearities used, the target, the type and strength of interventions, the noise tail behavior and whether parental contributions are multiplicative or additive. The simulation settings are described in Appendix~\ref{supp:sec:sim_settings} in detail. 

We apply all the 
conditional independence tests (CITs) 
that we have introduced in Section~\ref{subsec:condtest}, implemented with the following methods and tests as subroutines: 
\begin{center}
{\small
\begin{tabular}{p{20mm} p{125mm}}
CIT & Implementation \\
\hline
\abovespace\belowspace
(A)  & KCI without Gaussian process estimation \\
(B)(i) & RP w/ Fourier random features \\
(B)(ii) & RP w/ Nystr\"om random features and RBF kernel \\

(B)(iii) & RP w/ Nystr\"om random features and polynomial kernel (random degree) \\

(B)(iv) & RP w/ provided polynomial basis (random degree) \\

(C)  & Random forest and $\chi^2$-test  \\
(D)(i) &  GAM with F-Test \\
(D)(ii) & GAM with Wilcoxon test \\
(D)(iii) &  Random forest with F-Test \\
(D)(iv) &  Random forest with Wilcoxon test \\
(E)(i)  &  GAM with Kolmogorov-Smirnov test\\
(E)(ii) & GAM with Levene's test + Wilcoxon test \\
(E)(iii)  &   Random forest with Kolmogorov-Smirnov test \\
(E)(iv) &  Random forest with Levene's test + Wilcoxon test \\
(F) & Quantile regression forest with Fisher's exact test  \\
\hline
\end{tabular}
}
\end{center}

As a disclaimer we have to note that KCI is implemented without Gaussian process estimation. The KCI results shown below might improve if the latter is added to the algorithm. 

\paragraph{Baselines.} We compare against a number of baselines. Importantly, most of these methods contain various model misspecifications when applied in the considered problem setting. Therefore, they would not be suitable in practice. However, it is instructive to study the effect of the model misspecifications on performance. 
\begin{enumerate}
	\item The method of Causal Additive Models (CAM)
	 \citep{Buhlmann2014annals} 
identifies graph structure based on nonlinear additive noise models \citep{Peters2014}.
Here, we apply the method 
	 in the following way. We run CAM separately in each environment and output the intersection of the causal parents that were retrieved in each environment. 
	Note that the method's underlying assumption of Gaussian noise is violated. 
 \item We run the PC algorithm \citep{Spirtes2000} in two different variants. We consider a variable to be the parent of the target if a \textit{directed} edge between them is retrieved; we discard undirected edges. In the first variant of PC we consider, the environment variable is part of the input; conditional independence testing within the PC algorithm is performed with KCI, for unconditional independence testing we use HSIC \citep{Gretton2005, gretton_ind}, using the implementation from \cite{Pfister2017} (denoted with `PC(i)' in the figures). 
 In the second variant, we run the PC algorithm on the pooled data (ignoring the environment information), testing for zero partial correlations (denoted with `PC(ii)' in the figures). Here, the model misspecification is the assumed linearity of the structural equations.
\item We compare against linear ICP \citep{Peters2014} where the model misspecification is the assumed linearity of the structural equations.
\item We compare against LiNGAM \citep{Shimizu2006}, run on the pooled data without taking the environment information into account. Here, the model misspecifications are the assumed linearity of the structural equations and the i.i.d.\ assumption which does not hold.  
\item We also show the outcome of a random selection of the parents that adheres to the FWER-limit by selecting the empty set ($\hat{S}=\emptyset$) with probability $1-\alpha$ and setting $\hat{S}=\{k\}$ for $k$ randomly and uniformly picked from $\{1,\ldots,p\}\setminus k'$ with probability $\alpha$, where $k'$ is the index of the current target variable. The random selection is guaranteed to maintain FWER at or below $1-\alpha$.
\end{enumerate}
Thus, all considered baseline models in 1.\ -- 4.\ ---except for `PC(i)`---contain at least slight model misspecifications.


\begin{figure*}[!t]
\center
    \includegraphics[page = 1, width=0.8\textwidth, keepaspectratio=true, trim = {0 0 0 25}, clip]{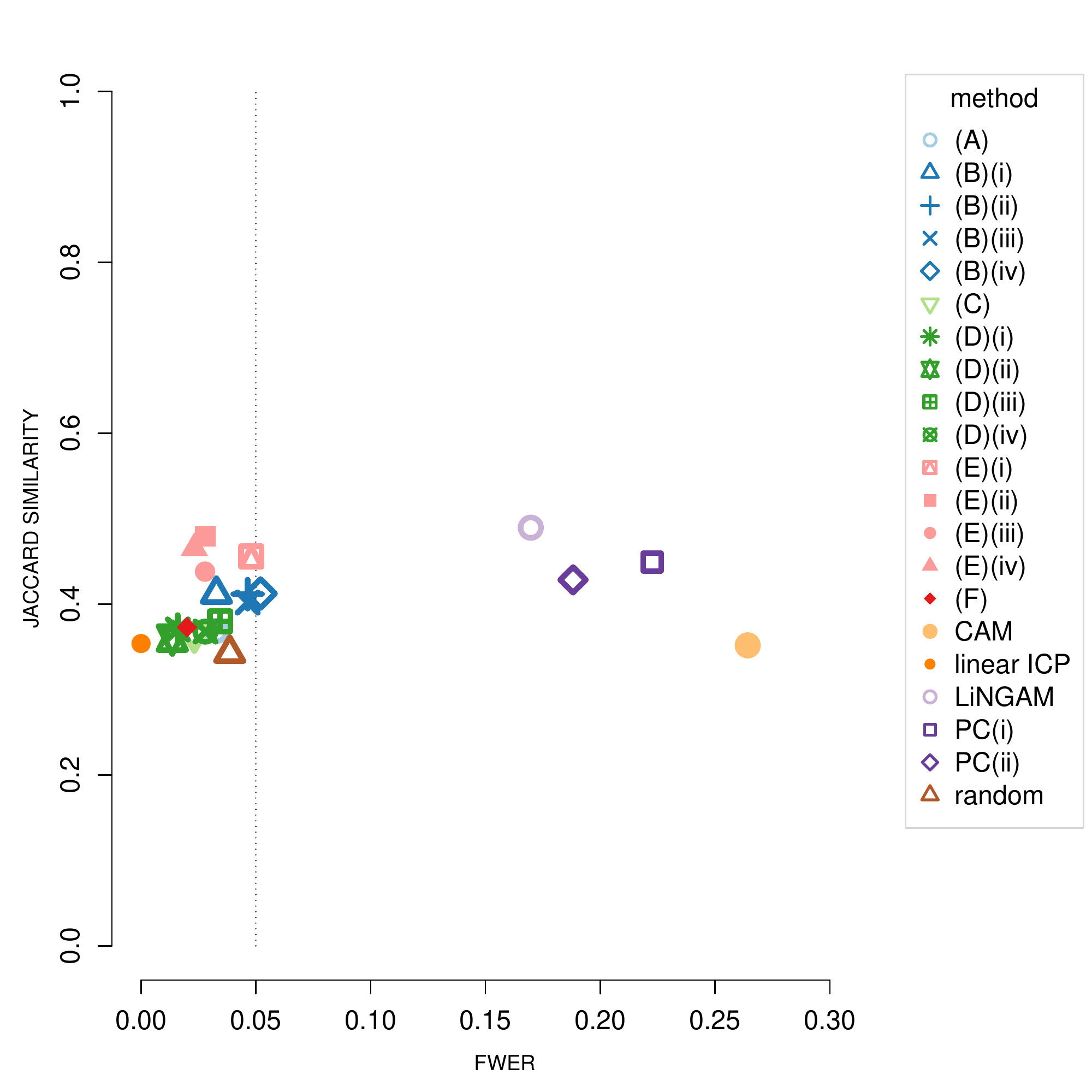}
\caption{\small  Average Jaccard similarity ($y$-axis) against average FWER ($x$-axis), stratified according to which conditional independence test (A) -- (F) or baseline method has been used. The nominal level is $\alpha=0.05$, illustrated by the vertical dotted line. The shown results are averaged over all target variables. Since the empty set is the correct solution for target variable 1 and 5, methods that mostly return the empty set (such as random or linear ICP) perform still quite well in terms of average Jaccard similarity. 
Since all variables are highly predictive for the target variable $Y$, see Figure~\ref{fig:dag_sim}, classical variable selection techniques as LASSO have a FWER that lies far beyond $\alpha$. 
Importantly, the considered baselines are not suitable for the considered problem setting due to various model misspecifications. We show their performance for comparison to illustrate the influence of these misspecifications.}\label{fig:performance_by_method}
\end{figure*}

\begin{figure*}[!t]
\center
    \includegraphics[page = 1, width=0.95\textwidth, keepaspectratio=true, trim = {0 0 0 15}, clip]{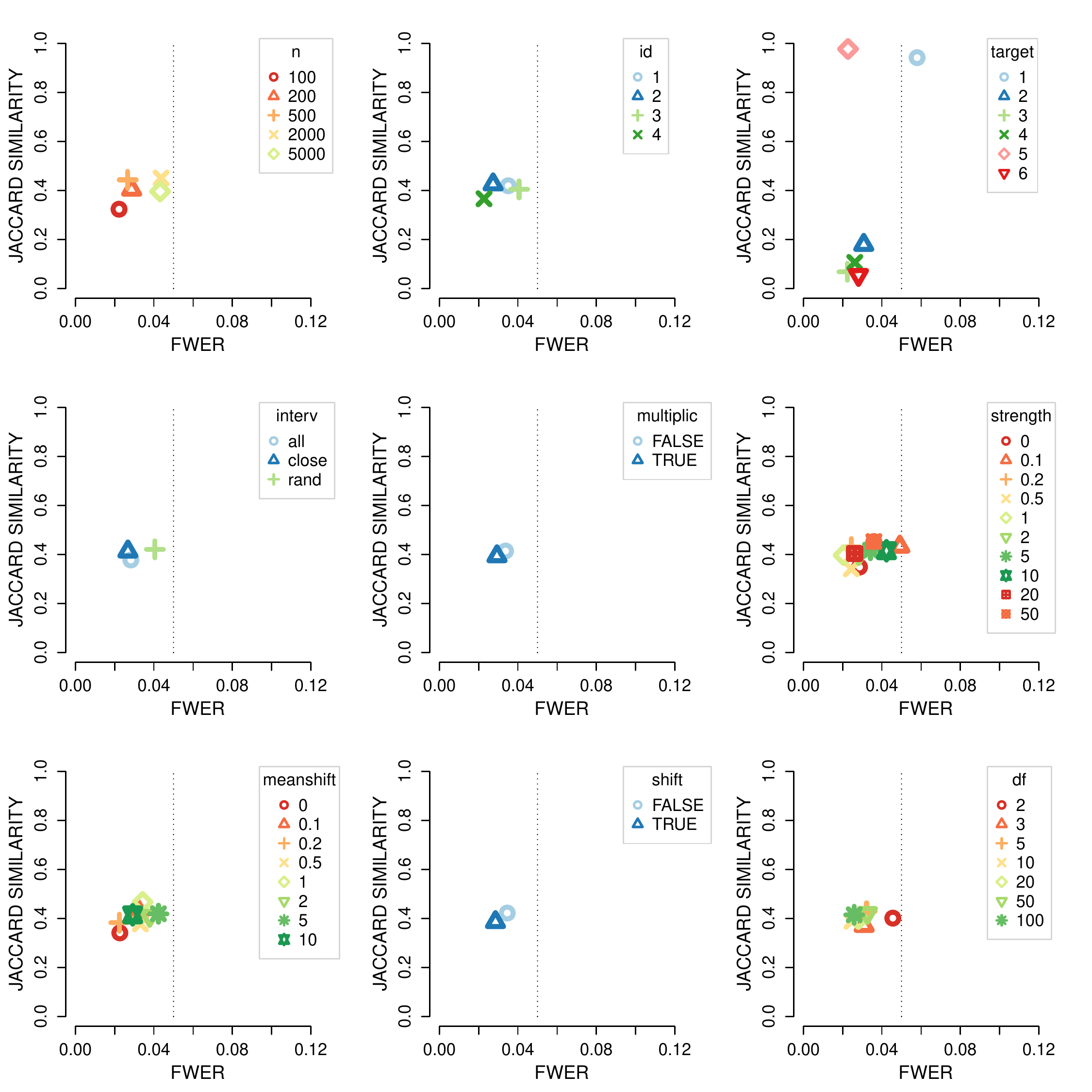}
\caption{\small Average Jaccard similarity over the conditional independence tests (A) -- (F) ($y$-axis) against average FWER ($x$-axis), stratified according to    various parameters (from top left to bottom right): sample size `n', type of nonlinearity `id', `target variable', intervention location `interv', multiplicative effects indicator `multiplic', `strength' of interventions, mean value of interventions `meanshift', shift intervention indicator `shift' and degrees of freedom for t-distributed noise `df'. For details, see the description in Appendix~\ref{supp:sec:sim_settings}. The FWER is within the nominal level in general for all conditional independence tests. The average Jaccard similarity is mostly determined by the target variable under consideration, see top right panel. }\label{fig:performance_by_param}
\end{figure*}

\begin{figure*}[!t]
\center
\subfloat[\small Target variable 2]{
	\includegraphics[page = 1, width=0.45\textwidth, keepaspectratio=true, trim = {0 0 0 0}, clip]{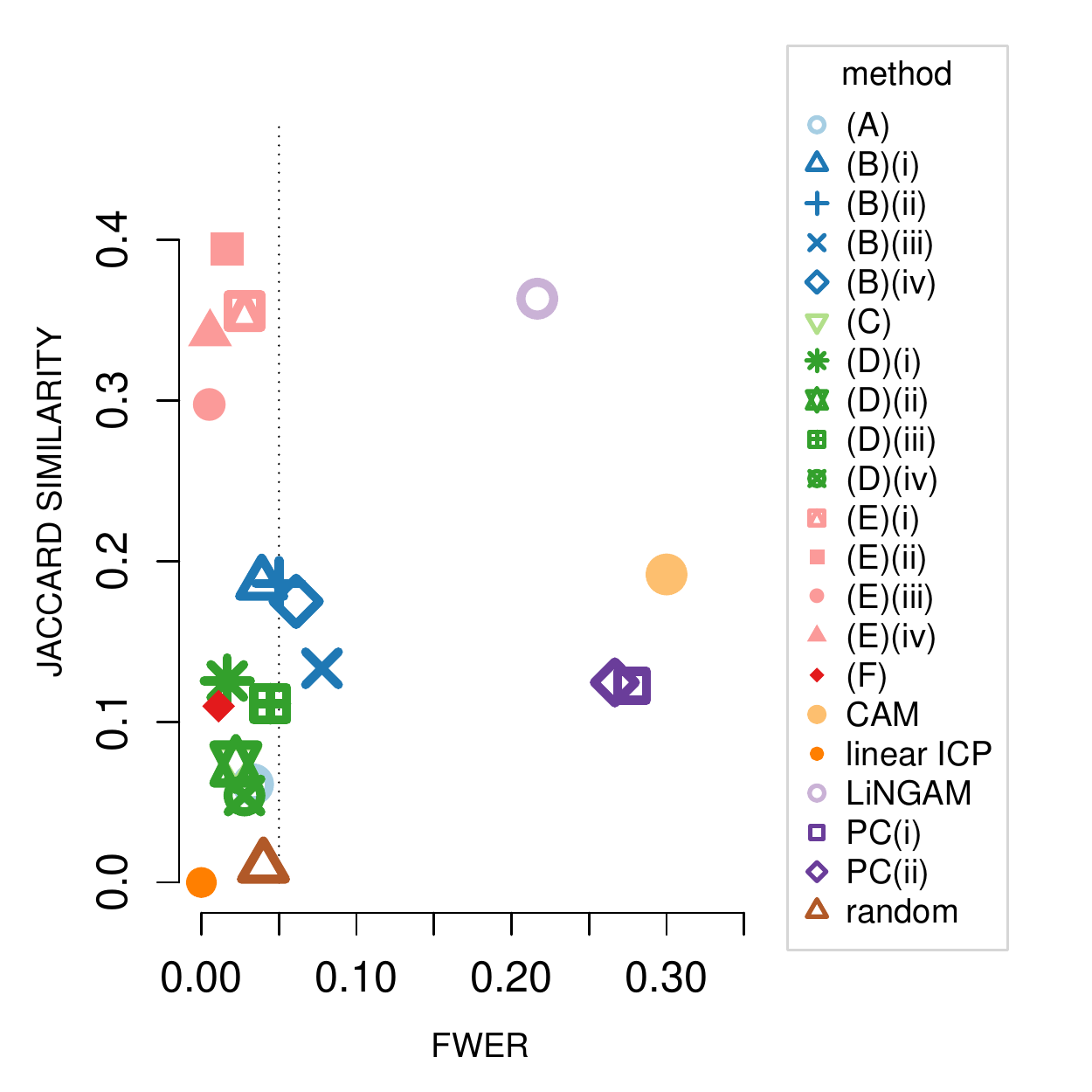}
	\label{fig:results_2}
}\subfloat[\small Target variable 3]{
	\includegraphics[page = 1, width=0.45\textwidth, keepaspectratio=true, trim = {0 0 0 0}, clip]{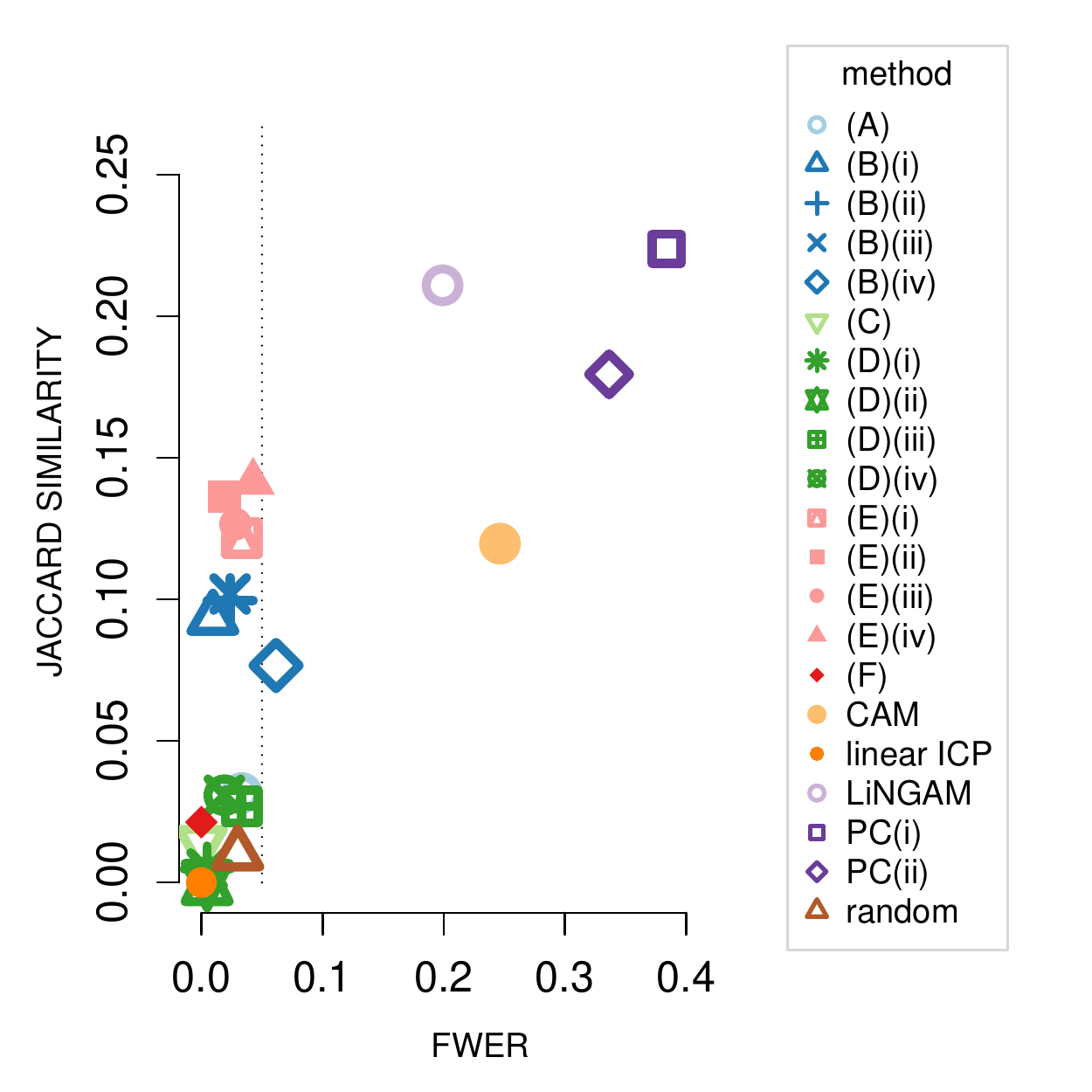}
	\label{fig:results_3}
}

\subfloat[\small Target variable 4]{ 
	\includegraphics[page = 1, width=0.45\textwidth, keepaspectratio=true, trim = {0 0 0 0}, clip]{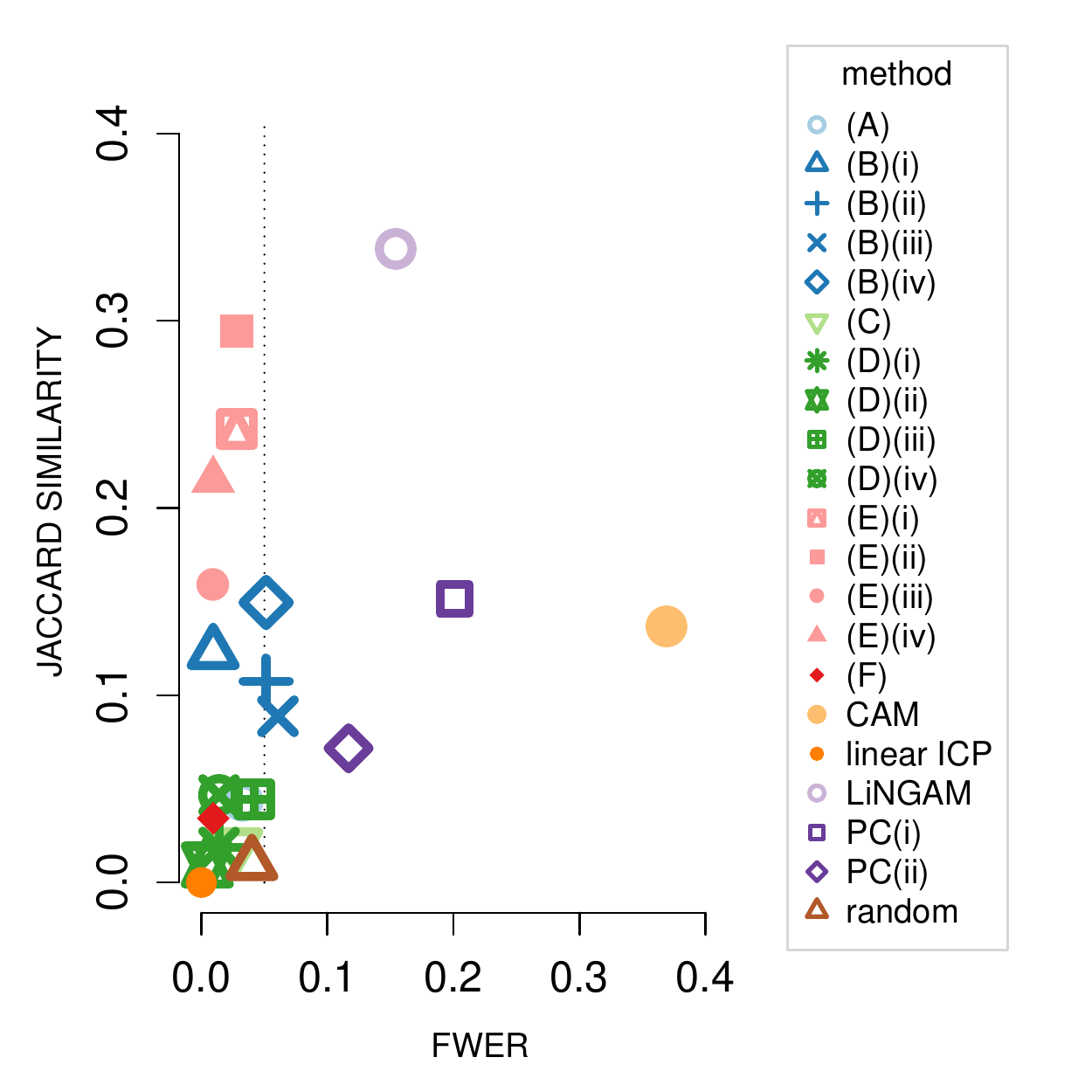}
	\label{fig:results_4}
}
\subfloat[\small Target variable 6]{
	\includegraphics[page = 1, width=0.45\textwidth, keepaspectratio=true, trim = {0 0 0 0}, clip]{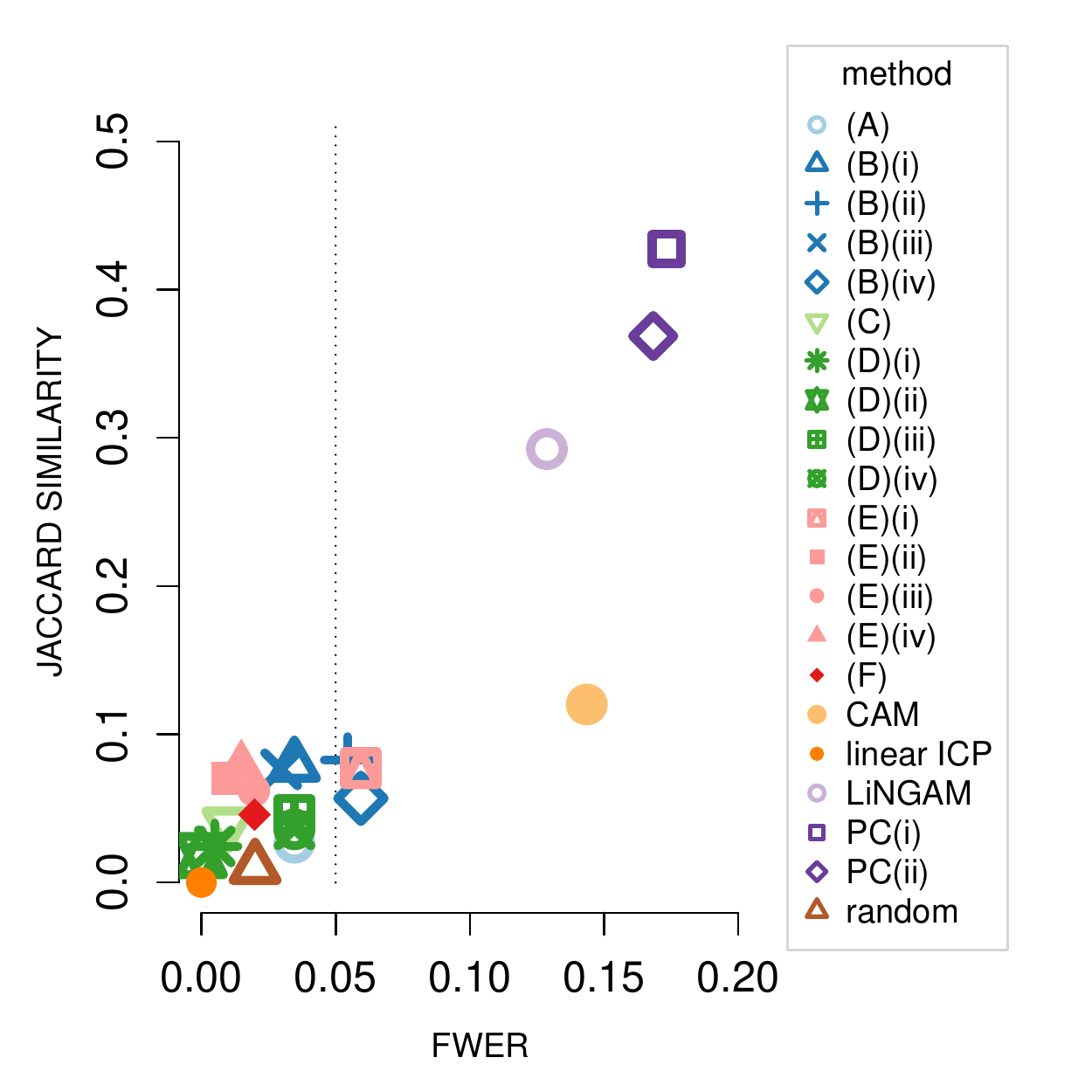}
	\label{fig:results_6}
}
 \caption{\small   The identical plot to Figure~\ref{fig:performance_by_method} separately for  target variables 2, 3, 4 and 6. For all target variables, method (E)(ii)---an invariant residual distribution test using GAM with Levene's test + Wilcoxon test---performs constantly as good or nearly as good as the optimal method among the considered tests. 
 }\label{fig:performance_by_method_variab}
 \end{figure*}

\paragraph{Metrics.} 
Error rates and power are measured in the following by
\begin{enumerate}[(i)]
\item Type I errors are measured by the {\bf family-wise error rate} (FWER), the probability of making one or more erroneous selections
\[ P\big( \hat{S} \nsubseteq S^{*}\big) .\]
\item Power is measured by the {\bf  Jaccard similarity}, the ratio between the size of the intersection and the size of the union of the estimated set $\hat{S}$ and the true set $S^{*}$. It is defined as~1 if both $S^{*}=\hat{S}=\emptyset$ and otherwise as 
\[  \frac{ \vert \hat{S} \cap S^{*}\vert }{\vert \hat{S} \cup S^{*}\vert} .\]
The Jaccard similarity is thus between 0 and 1 and the optimal value 1 is attained if and only if $\hat{S}=S^*$.
\end{enumerate}

\paragraph{Type-I-error rate of conditional independence tests.}
Figure~\ref{fig:performance_by_method} shows the average FWER on the $x$-axis
 (and the average Jaccard similarity on the $y$-axis) for all methods. The FWER is close but below the nominal FWER rate of $\alpha=0.05$ for all conditional independence tests, that is $P(\hat{S}\subseteq S^{*}) \ge 1-\alpha$.  
The same holds for the baselines linear ICP and random selection. Notably, the average Jaccard similarity of the random selection baseline is on average not much lower than for the other methods. The reason is mostly a large variation in average Jaccard similarity across the different target variables, as discussed further below and as will be evident from Figure~\ref{fig:performance_by_param} (top right plot). In fact, as can be seen from Figure~\ref{fig:performance_by_method_variab}, random guessing is much worse than the optimal methods on each target variable.
The FWER of the remaining baselines CAM, LiNGAM, PC(i) and PC(ii) lies well above $\alpha$. 

A caveat of the FWER control seen in Figure~\ref{fig:performance_by_method} is that while the FWER is maintained at the desired level, the test $H_{0,S^*}$ might be rejected more often than with probability $\alpha$. The error control rests on the fact that $H_{0,S^*}$ is accepted with probability higher than $1-\alpha$ (since the null is true for $S^*$). However, if a mistake is made and $H_{0,S^*}$ is falsely rejected, then we might still have $\hat{S}\subseteq S^*$  because 
either 
all other sets are rejected, too, in which case $\hat{S} = \emptyset$, or because
other sets (such as the empty set) are accepted and the intersection of all accepted sets is---by accident---again a subset of $S^*$.
In other words: some mistakes might cancel each other out but overall the FWER is very close to the nominal level, even if we  stratify according to sample size, target, type of nonlinearity and other parameters, as can be seen from Figure~\ref{fig:performance_by_param}.
 
\paragraph{Power.} Figures~\ref{fig:performance_by_method}  shows on the $y$-axis the average Jaccard similarity for all methods. The optimal value is 1  and is attained if and only if $\hat{S}=S^*$. A value 0 corresponds to disjoint sets $\hat{S}$ and $S^*$. The average Jaccard similarity is around 0.4 for most methods and not clearly dependent on the type I errors of the methods. Figure~\ref{fig:performance_by_param} shows the average FWER and Jaccard similarities stratified according to various parameters.

One of the most important determinants of success (or the most important)  is the target, that is the variable for which we would like to infer the causal parents; see top right panel in Figure~\ref{fig:performance_by_param}. 
Variables 1 and 5 as targets have a relatively high average Jaccard similarity when trying to recover the parental set. 
These two variables have an empty parental set ($S^*=\emptyset$) and the average Jaccard similarity thus always exceeds $1-\alpha$ if the level of the procedure is maintained as then $\hat{S}=\emptyset=S^*$ with probability at least $1-\alpha$ and the Jaccard similarity is~1 if both $\hat{S}$ and $S^*$ are empty. As testing for the true parental set corresponds to an unconditional independence test in this case, maintaining the level of the test procedure is much easier than for the other variables. 

Figure~\ref{fig:performance_by_method_variab} shows the same plot as Figure~\ref{fig:performance_by_method} for each of the more difficult target variables 2, 3, 4, and 6 separately. As can be seen from the graph in Figure~\ref{fig:dag_sim} and the detailed description of the simulations in Appendix~\ref{supp:sec:sim_settings},
the parents of target variable 3 are
difficult to estimate as the paths $1\rightarrow 2 \rightarrow 3$ and $1\rightarrow 3$ cancel each other exactly in the linear setting (and approximately for nonlinear data), thus creating a non-faithful distribution. The cancellation of effects holds true if interventions occur on variable 1 and not on variable 2. A local violation of faithfulness leaves type I error rate control intact but can hurt power as many other sets besides the true $S^*$ can get accepted, especially the empty set, thus yielding $\hat{S}=\emptyset$ when taking the intersection across all accepted sets to compute the estimate $\hat{S}$ in~\eqref{eq:Shat}. 
Variable 4, on the other hand, has only a single parent, namely $S^*=\{3\}$, and the recovery of the single parent is much easier, with average Jaccard similarity up to a third. Variable 6 finally again has average Jaccard similarity of up to around a tenth only. It does not suffer from a local violation of faithfulness as variable 3 but the size of the parental set is now three, which again hurts the power of the procedure, as often already a subset of the true parents will be accepted and hence $\hat{S}$ in~\eqref{eq:Shat} will not be equal to $S^*$ any longer but just a subset. For instance, when variable 5 is not intervened on in any environment it cannot be identified as a causal parent of variable 6, as it is then indistinguishable from the noise term. Similarly, in the linear setting, merely variable 3 can be identified as a parent of variable 6 if the interventions act on variables 1 and/or 2 only. 

The baselines LiNGAM and PC show a larger Jaccard similarity for target variables 3, 4 (only LiNGAM), and 6 at the price of large FWER values.

In Appendix~\ref{supp:add_results}, Figures~\ref{fig:performance_by_param2} -- \ref{fig:performance_by_param6} show the equivalent to Figure~\ref{fig:performance_by_param}, separately for target variables 2, 3, 4 and 6. 
For the sample size $n$, we observe that increasing it from 2000 to 5000 decreases power in case of target variables 4. This behavior can be explained by the fact that when testing $S^{*}$ in Eq.~\eqref{eq:H0S}, the null is rejected too often as the bias in the estimation performed as part of the conditional independence test yields deviations from the null that become significant with increasing sample size. 
For the nonlinearity, we find that the function $f_4(x) = \sin(2\pi x)$ is the most challenging one among the nonlinearities considered. It is associated with very low Jaccard similarity values for the target variables that do have parents. 
For the intervention type, it may seem surprising that `all' does not yield the largest power. A possible explanation is that intervening on all variables except for the target yields more similar intervention settings---the intervention targets do not differ between environments 2 and 3, even though the strength of the interventions is different. So more heterogeneity between the intervention environments, i.e.\ also having different intervention targets,  seems to improve performance in terms of Jaccard similarity.
Lastly, we see that power is often higher for additive parental contributions than for multiplicative ones. 

In summary, all tests (A) -- (F) seem to maintain the desired type I error, chosen here as the family-wise error rate, while the power varies considerably. 
An invariant residual distribution test using GAM with Levene's test and Wilcoxon test produces results here that are constantly as good or nearly as good as the optimal methods for a range of different settings. However, it is only applicable for categorical environmental variables. For continuous  environmental variables, the results suggest that the residual prediction test with random features might be a good choice. 


\section{Discussion and future work}
Causal structure learning with the invariance principle was proposed \cite{PetBuhMei15}. However, the assumption of linear models in \cite{PetBuhMei15}  is  unrealistic in many applications.  
In this work, we have shown how the framework can be extended to nonlinear and nonparametric  models by using suitable nonlinear and nonparametric conditional independence tests. The properties of these conditional independence tests are critically important for the power of the resulting causal discovery procedure. We evaluated many different test empirically in the given context and highlighted approaches that seem to work robustly in different settings. In particular we find that fitting a nonlinear model with pooled data  and then testing for differences between the residual distributions across environments results in desired coverage and high power if compared against a wide range of alternatives.

Our approach allowed us to model how several interventions may affect the total fertility rate of a country, using historical data about decline and rise of fertilities across different continents. In particular, we provided bounds on the average causal effect  under certain (hypothetical) interventions such as a reduction in child mortality rates. We showed that the causal prediction intervals for hold-out data have better coverage than various baseline methods. The importance of infant mortality rate and under-five mortality rate on fertility rates is highlighted, reconfirming previous studies that have shown or hypothesized these factors to be important \citep{hirschman1994fertility, raftery1995demand}. We stress that the results rely on causal sufficiency of the used variables, an assumption that can and should be debated for this particular example. 

We also  introduced the notion of `defining sets' in the causal discovery context  that helps in situations where the signal is weak or variables are highly correlated by returning sets of variables of which we know that at least one variable (but not necessarily all) in this set are causal for the target variable in question. 

Finally, we provide software in the \textsf{R} \citep{R-language} package  \texttt{nonlinearICP}. A collection of the discussed conditional independence tests are part of the package \texttt{CondIndTests}  and  are hopefully of independent interest. 

In applications where it is unclear whether the underlying models are linear or not, we suggest the following. While our proposed methods also hold the significance level if the underlying models are linear, we expect the linear version of ICP to have more power. Therefore, it is advisable to use the linear version of ICP if one has strong reasons to believe that the underlying model is indeed linear.
In practice, one might first apply ICP with linear models and apply a nonlinear version if, for example, all linear models are rejected. One would then need to correct for multiple testing by a factor of 2.

\bibliography{bibliography}

\section*{Acknowledgements}
We thank Jan Ernest and Adrian Raftery for helpful discussions and an AE and two referees for very helpful comments on an earlier version of the manuscript.

\appendix

\section{Time series bootstrap procedure}\label{supp:tsboots}
In the time series bootstrap procedure used to obtain the confidence bands $\hat{\mathcal{F}}$ in Section~\ref{subsec:prediction}, $B$ bootstrap samples of the response $Y$ are
generated by first fitting the model on all data points. We then use
the fitted values and residuals from this model. Each  bootstrap sample is generated by resampling the residuals of this fit block- and country-wise. 
In more detail, we define the block-length $\blocklen$ of residuals that should be sampled consecutively (we use $\blocklen=3$) and we sample a number of time points $t_{s_1}, \ldots, t_{s_k}$ from which the residuals are resampled. 
For a country $a$ and the first time point $t_1$, consider the fitted values at point $t_1$ and the fitted values for the $\blocklen-1$ consecutive observations. We then sample a country $b$ and add country $b$'s 
residuals from time points $t_{s_1}, t_{s_1+1}, \ldots t_{s_1+\blocklen-1}$ to the fitted values of country $a$ for the considered period $t_1, \ldots, t_{\blocklen}$. We then proceed with the next $\blocklen$ consecutive fitted values for country $a$ and add country $b$'s 
residuals from observations $t_{s_2}, t_{s_2+1}, \ldots
t_{s_2+\blocklen-1}$, until all fitted values of country $a$ are
covered. This procedure is applied  to each country.
Finally, to obtain the confidence intervals, we fit the model on each of the $B$ bootstrap samples $(Y^b,X)$, consisting of the response $Y^b$ generated from the fitted values and the resampled residuals, and the observations $X$ which have not been modified. 

\section{Conditional independence tests}\label{supp:sec:condtest_details}

For completeness, we first restate the generic method for Invariant Causal Prediction from \cite{PetBuhMei15}:
\begin{algorithm}[H]
\caption{Generic method for Invariant Causal Prediction} \label{alg:icp_generic}
\algorithmicrequire\; i.i.d.\ sample of ($Y$, $X$, $E$), $\alpha$ 

\begin{algorithmic}[1]
\FOR{\textbf{each} $S \subseteq \{1, \ldots, p \}$}
    \STATE Test whether \hypnonlinICP holds at level $\alpha$. 
\ENDFOR  
	\STATE Set $\hat{S} := \bigcap_{S:H_{0,S} \text{ not rejected}} S $ 
\end{algorithmic}
  \algorithmicensure\; $\hat{S}$
\end{algorithm}

The conditional independence tests discussed in this work can be used to perform the test in Step 2 of Algorithm~\ref{alg:icp_generic}. Therefore, the inputs to these tests consist of an i.i.d.\ sample of $(Y, X_S, E)$ and $\alpha$ where $X_S$ contains the variables corresponding to $S \subseteq \{1, \ldots, p \}$, i.e.\ the subset to be tested. 
Additionally, some test specific parameters might need to be specified. The return value of the tests is the respective test's decision about \hypnonlinICP.

For most tests, $\envvar \in \mathbb{R}^d$ can be either discrete or continuous. As all empirical results in this work use an environment variable that is discrete and one-dimensional, the descriptions below focus on this setting. We then denote the index set of different  environments with $\E$. 
We will comment on the required changes for the continuous and higher-dimensional case in the respective sections. Whenever applying the test for environmental variables $\envvar \in \mathbb{R}^d$ with $d > 1$ is infeasible with the method, each test can be applied separately for each variable in $\envvar$. The overall $p$-value is obtained by multiplying the minimum of the individual $p$-values by $d$, i.e.\ by applying a Bonferroni correction for the number of environmental variables. When applying the function \texttt{CondIndTest()} from the \textsf{R} package \texttt{CondIndTests} with a conditional independence test that does not support a multidimensional environment variable, the described Bonferroni correction is applied. 

\subsection{Kernel conditional independence test}\label{supp:ssec:kernel}
\paragraph{Setting and assumptions.} We use  the kernel conditional independence test proposed in \cite{Zhang11}. When $\envvar$ is discrete, we use a delta kernel for  $\envvar$, and otherwise an RBF kernel.  The test is also applicable when $\envvar$ contains more than one environmental variable as the inputs can be sets of random variables.

\subsection{Residual Prediction test}\label{supp:ssec:rptest}

\paragraph{Setting and assumptions.} We do not expect this test to have the correct level when the noise in Eq.~\eqref{eq:Y} is not additive. The described procedure does not need to be modified for higher-dimensional and/or continuous environmental variables $\envvar$.

We consider a version of a Residual Prediction test as proposed in \cite{Shah15} to determine whether \hypnonlinICP holds at level $\alpha$ for a particular set of variables $S$. The main idea is to find a suitable basis expansion of $f$ 
that allows us to regress $Y$ on $X_S$ by reverting back to the linear case. Given an appropriate basis expansion, the \emph{scaled} ordinary least squares residuals can then be tested for possible remaining nonlinear dependencies between the scaled residuals and $(E, X_S)$.
The scaling ensures that the resulting test statistic is not a function of the noise variance. Under the null, the scaled residuals are expected to behave roughly like the noise term. In other words, there should be no dependence between the scaled residuals and the environmental variables and $X_S$, so there should be no signal left in the residuals that could be fitted by a nonparametric method like a random forest using $\envvar$ and $X_S$ as predictors. This necessitates to make an assumption on the noise distribution $F_\varepsilon$, e.g.\ $\varepsilon \sim \mathcal{N}(0, 1)$.  

In order to generalize the method to settings where an appropriate basis expansion of $f$ is unknown, we look at ways to find such a suitable basis expansion automatically by using random features \citep{Williams2000, Rahimi2007}.

\begin{algorithm}[H]
\caption{Residual Prediction tests applied to nonlinear ICP  \label{alg:RP_test}}
	\algorithmicrequire\; i.i.d.\ sample of ($Y$, $X_S$, $\envvar$), $\alpha$, $F_\varepsilon$, $B$, a subroutine to compute the basis functions $h_m(\cdot) \text{ for } m = 1, \ldots, M$ 
	
\begin{algorithmic}[1]
        \STATE Compute the non-linear transformations 
$
h_m(\Xs),  m = 1, \ldots, M
$	and create the design matrix $\mathbf{H}_{\Xs} \in \mathbb{R}^{n \times M}$ comprising these $M$ nonlinear features. 
	\STATE Regress $Y$ on $\mathbf{H}_{\Xs}$ with ordinary least squares.
	\STATE  Predict (a function of) the scaled residuals with the environment variable  $\envvar$ and $\Xs$. 
	\STATE Compute a statistic for the prediction accuracy to be used as test statistic.
	\FOR{$b$ from $1$ to $B$}
	\STATE Simulate one sample of size $n$ from the assumed noise distribution $F_\varepsilon$.
	\STATE Predict (a function of) these simulated values after rescaling with the environment variable  $\envvar$ and $\Xs$. 
	\STATE Compute a statistic for the prediction accuracy.
	\ENDFOR
	\STATE The $B$ simulated values for prediction accuracy yield the empirical null distribution from which the $p$-value is obtained.
\end{algorithmic}
  \algorithmicensure\; Decision about \hypnonlinICP
\end{algorithm} 

\paragraph{Step 1.} The choice of $h_m(\Xs), m = 1, \ldots, M$ can be based on domain knowledge, e.g.\ when the nonlinearity in Eq.~\eqref{eq:Y} is known to be a polynomial of a given order. If such domain knowledge is not available, the linear basis expansion can be approximated by random features, e.g.\ using the Nystr\"om method or by random Fourier features. For these methods, the kernel function needs to be chosen as well as the kernel parameters and the number of random features to be generated. 
\paragraph{Step 3.} For instance, a random forest can be used for the estimation. If the residuals only differ in the second moments,
predicting the expectation of the residuals is not sufficient as the predictors $\envvar$ have no discriminative power for this task. In such a setting, the absolute value of the residuals can be predicted to exploit the heterogeneity in the second moments across environments.

\paragraph{Step 4.} For instance, the mean squared error can be used here. 
	 
\paragraph{Step 5.} If the error term is non-Gaussian, the appropriate distribution can be used at this stage to accommodate non-Gaussianity of the noise.

\paragraph{Parameter settings used in simulations.}
In the simulations, we use $B = 250$ and $\varepsilon \sim \mathcal{N}(0, 1)$. 
In step~1, we consider the following options: (a) Fourier random features (approach (B)(i) in Section~\ref{sec:experiments}), (b) Nystr\"om random features and RBF kernel ((B)(ii)), (c) Nystr\"om random features and polynomial kernel of random degree ((B)(iii)), (d)  polynomial basis of random degree ((B)(iv)). 
The number of random features in (c) and (d) is chosen to be  $\lceil n/4 \rceil$.
In step~7, we predict the mean as well as the absolute value of the residuals and aggregate the results using a Bonferroni correction. 

\subsection{Invariant environment prediction}\label{supp:ssec:envpred}

\paragraph{Setting and assumptions.} 
 The described procedure does not need to be modified for continuous environmental variables $\envvar$. For higher-dimensional $\envvar$ the test would need to be applied for each variable separately and the resulting $p$-values would need to be aggregated with a Bonferroni correction.

\begin{algorithm}[H]
\caption{Invariant environment prediction for nonlinear ICP \label{alg:targetE}}
	\algorithmicrequire\; i.i.d.\ sample of ($Y$, $X_S$, $\envvar$), $\alpha$, subroutine for test in step~5. 
	
\begin{algorithmic}[1]
\STATE Split the sample into training and test set. 
\STATE Use the training set to train a model to predict $\envvar$ with $(Y, X_S)$ as predictors.
\STATE Use the training set to train a model to predict $\envvar$ with $X_S$ as predictors. 
\STATE For both fits, compute the prediction accuracy on the test set.
\STATE Use a one-sided test at the significance level $\alpha$ to assess whether the prediction accuracy of the fit using $(Y, X_S)$ as predictors is larger than the prediction accuracy of the fit using only $X_S$ as predictors.
\end{algorithmic}
  \algorithmicensure\; Decision about \hypnonlinICP
\end{algorithm} 

\paragraph{Step 3.} When a random forest is used to predict the environment variable, one can also use $X_S$ and a permutation of $Y$ as predictors to ensure the random forest fits are based on the same number of predictor variables. As the number of variables considered for each split in the random forest estimation procedure is a function of the total number of predictor variables, this helps to mitigate differences between the prediction accuracies that are only due to artefacts of the estimation procedure. This is especially relevant for small sets $S$.

\paragraph{Step 5.} For instance, a $\chi^2$ test can be used here.  
If the null is true and we find the optimal model in both cases, then the out-of-sample performance of both models is statistically indistinguishable as $Y$ is independent of $\envvar$ given $X_S$. If the null is not true, we expect the model containing the response to perform better as $Y$ contains additional information in this case (since $Y$ is not independent of $\envvar$ given $X_S$).

\paragraph{Parameter settings used in simulations.}
In step~1, we use $2/3$ of the data points for training and $1/3$ for testing. In step~3, we use a random forest to predict the environment variable and use $X_S$ and a permutation of $Y$ as predictors. In step~4, we use the $\chi^2$ test implemented in \texttt{prop.test()} \citep{Wilson1927} from the \texttt{stats} package in \textsf{R}. 

\subsection{Invariant target prediction}\label{supp:ssec:targetpred}

\paragraph{Setting and assumptions.} 
The described procedure does not need to be modified for continuous and/or higher-dimensional environmental variables $\envvar$. 

\begin{algorithm}[H]
\caption{Invariant target prediction for nonlinear ICP \label{alg:targetY}}
\algorithmicrequire\; i.i.d.\ sample of ($Y$, $X_S$, $\envvar$), $\alpha$, subroutine for test in step~5.

\begin{algorithmic}[1]
\STATE Split the sample into training and test set. 
\STATE Use the training set to train a model to predict $Y$ with $(X_S, \envvar)$ as predictors.
\STATE Use the training set to train a model to predict $Y$ with $X_S$ as predictors. 
\STATE For both fits, compute the prediction accuracy on the test set.
\STATE Use a one-sided test at the significance level $\alpha$ to assess whether the prediction accuracy of the fit using $(X_S, \envvar)$ as predictors is larger than the prediction accuracy of the fit using only $X_S$ as predictors.
\end{algorithmic}
  \algorithmicensure\; Decision about \hypnonlinICP
\end{algorithm} 

\paragraph{Step 3.} When a random forest is used, one can also use $X_S$ and a permutation of $\envvar$ as predictors to ensure the random forest fit is based on the same number of predictor variables. As the number of variables considered for each split in the random forest estimation procedure is a function of the total number of predictor variables, this helps to mitigate differences between the prediction accuracies that are only due to artefacts of the estimation procedure. This is especially relevant for small sets $S$. As an alternative to using a random forest, one could use GAMs as the estimation procedure, implying the implicit assumption that the components in $f(X)$ in Eq.~\eqref{eq:Y} are additive. 

\paragraph{Step 5.} For instance, an F-test can be used here. Another option is a Wilcoxon test using the difference between the absolute residuals. If the null is true and we find the optimal model in both cases, then the out-of-sample performance of both models is statistically indistinguishable as $Y$ is independent of $\envvar$ given $X_S$. If the null is not true, we expect the model additionally containing $\envvar$ to perform better as $\envvar$ contains additional information in this case (since $Y$ is not independent of $\envvar$ given $X_S$).

\paragraph{Parameter settings used in simulations.}
In step~1, we use $2/3$ of the data points for training and $1/3$ for testing. In step~3, to predict $Y$ we use a GAM or a random forest. In step~5, we use an F-test or a Wilcoxon test (\texttt{wilcox.test()} from the \texttt{stats} package in \textsf{R}). These combinations yield approaches (D)(i) -- (iv) in Section~\ref{sec:experiments}. When using a random forest in step~3, we use $X_S$ and a permutation of $\envvar$ as predictors. 

\subsection{Invariant residual distribution test}\label{supp:ssec:residdis}
\paragraph{Setting and assumptions.} We do not expect this test to have the correct level when the noise in Eq.~\eqref{eq:Y} is not additive. It is only applicable to discrete environmental variables. For higher-dimensional $\envvar$ the test would need to be applied for each variable separately and the resulting $p$-values would need to be aggregated with a Bonferroni correction.

\begin{algorithm}[H]
\caption{Invariant residual distribution test for nonlinear ICP \label{alg:errorDis}}
	\algorithmicrequire\; i.i.d.\ sample of ($Y$, $X_S$, $\envvar$), $\alpha$, subroutine for test in step 4. 
\begin{algorithmic}[1]
\STATE Pool the data from all environments and fit a model to predict $Y$ with $X_S$. 
\STATE Initialize $\pval\leftarrow 1$, $t \leftarrow 0$.
\FOR{\textbf{each} $e \in \E$}
	
    \STATE Use a two-sample test to assess whether the residuals of samples from environment $e$ have the same distribution as the residuals of samples from environments in the index set $\E'$ where $\E' = \E \setminus \{e\}$, yielding the $p$-value $\pval_{e}$.
    \STATE $t \leftarrow t + 1$
    \STATE $\pval\leftarrow \min(\pval, \pval_{e})$.
    \IF{ $\vert \E \vert = 2$} 
    \STATE break
    \ENDIF
\ENDFOR
	\STATE Apply a Bonferroni correction for the number of performed tests $t$: $\pval\leftarrow t \cdot \pval$.
\end{algorithmic}
  \algorithmicensure\; Decision about \hypnonlinICP
\end{algorithm} 

\paragraph{Step 1.} For instance, one could use a random forest or a GAM as the estimation procedure. The latter implicitly assumes that the components in $f$ in Eq.~\eqref{eq:Y} are additive.

\paragraph{Step 4.} For instance, a nonparametric test such as Kolmogorov-Smirnov can be used here. Alternatively, we can limit the test to assess equality of first and second moments by first using a Wilcoxon test for the expectation with an one-vs-all scheme as described in the algorithm. Subsequently, Levene's test for homogeneity of variance across groups can be used to test for equality of the second moments of the residual distributions. In this case, the final $p$-value would be twice the minimum of (a) the Bonferroni-corrected $p$-value from the one-vs-all Wilcoxon test and (b) the $p$-value from Levene's test.

\paragraph{Parameter settings used in simulations.}
In step~1, we use a GAM or a random forest. In step~4, we use both approaches described above, using (a)  \texttt{ks.test()} from the \texttt{stats} package in \textsf{R} \citep{conover:1971} and (b) \texttt{wilcox.test()} and \texttt{levene.test()} (the latter being contained in the \texttt{lawstat} package in \textsf{R} \citep{Levene1960, levene.test}). These combinations yield approaches (E)(i) -- (iv) in Section~\ref{sec:experiments}. 

\subsection{Invariant conditional quantile prediction}\label{supp:ssec:invquant}
\paragraph{Setting and assumptions.}  
For continuous and/or higher-dimensional environmental variables $\envvar$ the test described in Steps 4 -- 11 which assesses whether $\text{Exceedance} \independent \envvar$ would need to be modified according to the structure of $\envvar$.
\begin{algorithm}[H]
\caption{Invariant conditional quantile prediction for nonlinear ICP} \label{alg:inva_quant} 
	\algorithmicrequire\; i.i.d.\ sample of ($Y$, $X_S$, $\envvar$), $\alpha$, set of quantiles $\mathcal{B}$, subroutine for test in step~7. 
\begin{algorithmic}[1]
\STATE Initialize $\pval \leftarrow 1$, $t \leftarrow 0$. 
\FOR{\textbf{each} $\beta \in \mathcal{B}$}
\STATE Predict $1-\beta$ quantile $Q_{1-\beta}(x)$ of $Y \vert {X_S = x}$.
\FOR{\textbf{each} $e \in \E$} 
\STATE $ \text{Define one-vs-all environment } I = \mathbbm{1}_{\{ \envvar = e \}}$ 
\STATE $\text{Define exceedance } E_{1-\beta} = \mathbbm{1}_{\{ Y > \hat{Q}_{1-\beta}(x) \}} $
\STATE  Test whether $ E_{1-\beta}$ is independent of $I$: $\pval_{e, \beta} \leftarrow \texttt{StatTest}(E_{1-\beta}, I, \alpha)$
\STATE  $t \leftarrow t + 1$, $\pval \leftarrow \min(\pval, \pval_{e, \beta})$
\IF{ $\vert \E \vert = 2$} 
\STATE break
\ENDIF
\ENDFOR
\ENDFOR
	\STATE  Apply a Bonferroni correction for the number of performed tests $t$: $\pval \leftarrow t \cdot \pval$
\end{algorithmic}
  \algorithmicensure\; Decision about \hypnonlinICP \end{algorithm}

\paragraph{Step 3.} For instance, a Quantile Regression Forest \citep{Meinshausen06quantileregression} can be used here. 

\paragraph{Step 7.} For instance, Fisher's exact test can be used here by computing the $2 \times 2$ contingency table of the exceedance of the residuals for the quantile $1-\beta$ for $I = 0$ and for $I = 1$.

\paragraph{Parameter settings used in simulations.}
In step~3, we use a quantile regression forest for $\mathcal{B} = \{0.1, 0.5, 0.9\}$. In step~7, we use \texttt{fisher.test()} from the \texttt{stats} package in \textsf{R}.

\subsection{Overview of conditional independence tests in \texttt{CondIndTests} package}\label{supp:ssec:package}
The described conditional independence tests are available in the \textsf{R} package \texttt{CondIndTests}. A wrapper function \texttt{CondIndTest()} is provided which takes the respective test as the argument \texttt{method}. The  package supports the estimation procedures, subroutines and statistical tests shown in Table~\ref{tab:package}. The column $\envvar$ indicates whether the environmental variables can be discrete ('D'), continuous ('C'), or both; the column $d$ shows the supported dimensionality of $\envvar$.

As described at the beginning of Appendix~\ref{supp:sec:condtest_details}, a Bonferroni correction is applied when calling the function \texttt{CondIndTest()} with a conditional independence test that does not support a multidimensional environment variable. Similarly, a Bonferroni correction is applied when the first input argument \texttt{Y} to the respective test is multidimensional and if the specified test does not support this internally. 

\renewcommand{\baselinestretch}{1.15}
\begin{table}[!htbp]
\caption{Overview of implemented test combinations in \texttt{CondIndTests} package}
\label{tab:package}
\begin{center}
\begin{tabular}{p{10mm} >{\raggedright}p{55mm} >{\raggedright}p{55mm} >{\raggedright}p{10mm} p{10mm}}
\hline
\abovespace\belowspace
{\sc CIT} & \textsf{R} function name/{\sc Method} & {\sc Test} & $\envvar$ & $d$ \\
\hline
\abovespace\belowspace
(A)  & \texttt{KCI()} \newline KCI (\small without GP support) & $\quad$ \newline -- & D/C & $\geq 1$ \\
\hline
\abovespace\belowspace
(B)  & \texttt{ResidualPredictionTest()} \newline Residual prediction test with 
\newline -- Nystr\"om random features (RBF and polynomial kernel)
\newline -- Fourier random features
\newline -- fixed basis expansion & $\quad$ \newline --  & D/C & $\geq 1$ \\
\hline
\abovespace\belowspace
(C)  & \texttt{InvariantEnvironmentPrediction()} \newline Random forest & $\quad$ \newline $\chi^2$ test (\texttt{prop.test()}) \newline Wilcoxon test (\texttt{wilcox.test()}) & D/C & $1$ \\
\hline
\abovespace\belowspace
(D)  & \texttt{InvariantTargetPrediction()} \newline Random forest \newline GAM & $\quad$\newline F-Test \newline Wilcoxon test (\texttt{wilcox.test()}) & D/C & $\geq 1$ \\
\hline
\abovespace
\belowspace
(E) & \texttt{InvariantResidualDistributionTest()}\newline Random forest \newline GAM & $\quad$ \newline Kolmogorov-Smirnov (\texttt{ks.test()}) \newline Levene's test + Wilcoxon test (\texttt{levene.test(), wilcox.test()}) & D & $ 1$ \\
\hline
\abovespace
\belowspace
(F) & \texttt{InvariantConditionalQuantilePrediction()}
 \newline Quantile regression forest & $\quad$ \newline Fisher's exact test (\texttt{fisher.test()}) & D & $1$ \\
\hline
\end{tabular}
\end{center}
\end{table}
\renewcommand{\baselinestretch}{1}

\section{Experimental settings for numerical studies}\label{supp:sec:sim_settings}

For each simulation, we compare the performance of all methods and conditional independence tests  while choosing the following parameters randomly (but keeping them constant for one simulation):
In total, there are 27478 simulations from 1240 distinct settings that are evaluated for each of the 22 considered methods.

\begin{enumerate}[(i)]
\item {\bf Sample size.} Sample size `n' is chosen randomly in the set $\{100,200,500,2000,5000\}$.
\item {\bf Target variable.} We sample one of the variables in the graph in Figure~\ref{fig:dag_sim} at random as a target variable (variable `target' is chosen uniformly from $\{1,2,\ldots,6\}$ in other words).
\item {\bf Tail behavior of the noise.} The noise $\eta_k$ for $k=1,\ldots,6$ is sampled from a t-distribution and the degrees of freedom are chosen at random from $\mathrm{df} \in \{2,3,5,10,20,50,100\}$, where the latter is very close to a Gaussian distribution.
\item {\bf Multiplicative or additive effects.}  For each simulation setting, we determine whether $g_k(\cdot)$ has additive or multiplicative components. We sample additive components of the form $g_{k}(Z_{\text{pa}_k}) = \sum_{j \in  pa_k } f(\epsilon_{j,k} \cdot Z_j)$  ($\mathit{multiplic}=\mathrm{FALSE}$) and multiplicative components of the form $g_{k}(Z_{\text{pa}_k}) = \prod_{j \in  pa_k } f(\epsilon_{j,k} \cdot Z_j)$ ($\mathit{multiplic}=\mathrm{TRUE}$) with equal probability, where the signs $\epsilon_{j,k}\in \{-1,1\}$ are as shown in Figure~\ref{fig:dag_sim} along the relevant arrows.
\item {\bf Shift- or do-Interventions.} The variable `shift' is set with equal probability to either $\mathrm{TRUE}$ (shift-interventions) or $\mathrm{FALSE}$ (do-interventions). For do-interventions we replace the structural equation of the intervened variable $k\in \{1,\ldots,q\}$ by 
\[ Z_k \leftarrow e_k ,\]
where $e_k$ is the randomly chosen intervention value which is sampled i.i.d\ for each observation as described under (vi). For shift-interventions, the value $e_k$ is added as
\[ Z_k \leftarrow g_{k}(Z_{\text{pa}_k}) + \eta_k+e_k.\]
See for example Section~5 of \cite{PetBuhMei15} for a more detailed discussion of shift interventions. 
\item {\bf Strength of interventions} 
The intervention values $e_k$ are chosen independently for all variables from a t-distribution with `df' degrees of freedom, shifted by a constant `meanshift' (chosen uniformly at random in $\{0,0.1,0.2,0.5,1,2,5,10\}$, and scaled by a factor `strength', chosen uniformly at random in $\{0,0.1,0.2,0.5,1,2,5,10\}$).
\item {\bf Non-linearities.} 
For the functions $f=f_{\mathrm{id}}$ we consider the following four nonlinear functions, where the index `id'  is sampled uniformly from $\{1,2,3,4\}$ and the same nonlinearity is used throughout the graph:
\begin{align*} f_1(x) &= x, \\ 
f_2(x) &= \max\{0,x\}, \\ 
f_3(x) &= \mbox{sign}(x) \cdot \sqrt{ |x|}, \\ 
f_4(x) &= \sin(2\pi x), 
\end{align*}
\item {\bf Location of interventions.}  Each sample is independently assigned into one environment in $\E=\{1,2,3\}$, where $\{1\}$ corresponds to observational data, that is all samples in environment $\{1\}$ are sampled as observational data, where samples in environments $\{2,3\}$ are intervention data.  The intervention targets and strengths for  samples in environment $\{2\}$ are  drawn  as per the description below and kept constant for all samples in environment $\{2\}$ and then analogously for environment $\{3\}$, where intervention targets are drawn independently and identically to environment $\{2\}$.  The variable `interv' is set uniformly at random to one of the values $\{$`all',`rand',`close'$\}$. If it is equal to `all', then interventions in environments $\{2,3\}$ occur at all variables except for the target variable. If it is equal to `rand', then interventions occur at one ancestor, chosen uniformly at random, and one descendant of the target variable, chosen again uniformly at random  (the empty set is chosen in case there are no ancestors or descendants). Finally, if it is equal to `close', interventions occur at one parent, chosen uniformly at random, and one child of the target variable, chosen again uniformly at random (and again no interventions occur if these sets are empty).

\end{enumerate}

\FloatBarrier
\section{Additional experimental results}\label{supp:add_results}

\begin{figure*}[!t]
\center
    \includegraphics[page = 1, width=0.95\textwidth, keepaspectratio=true, trim = {0 0 0 0}, clip]{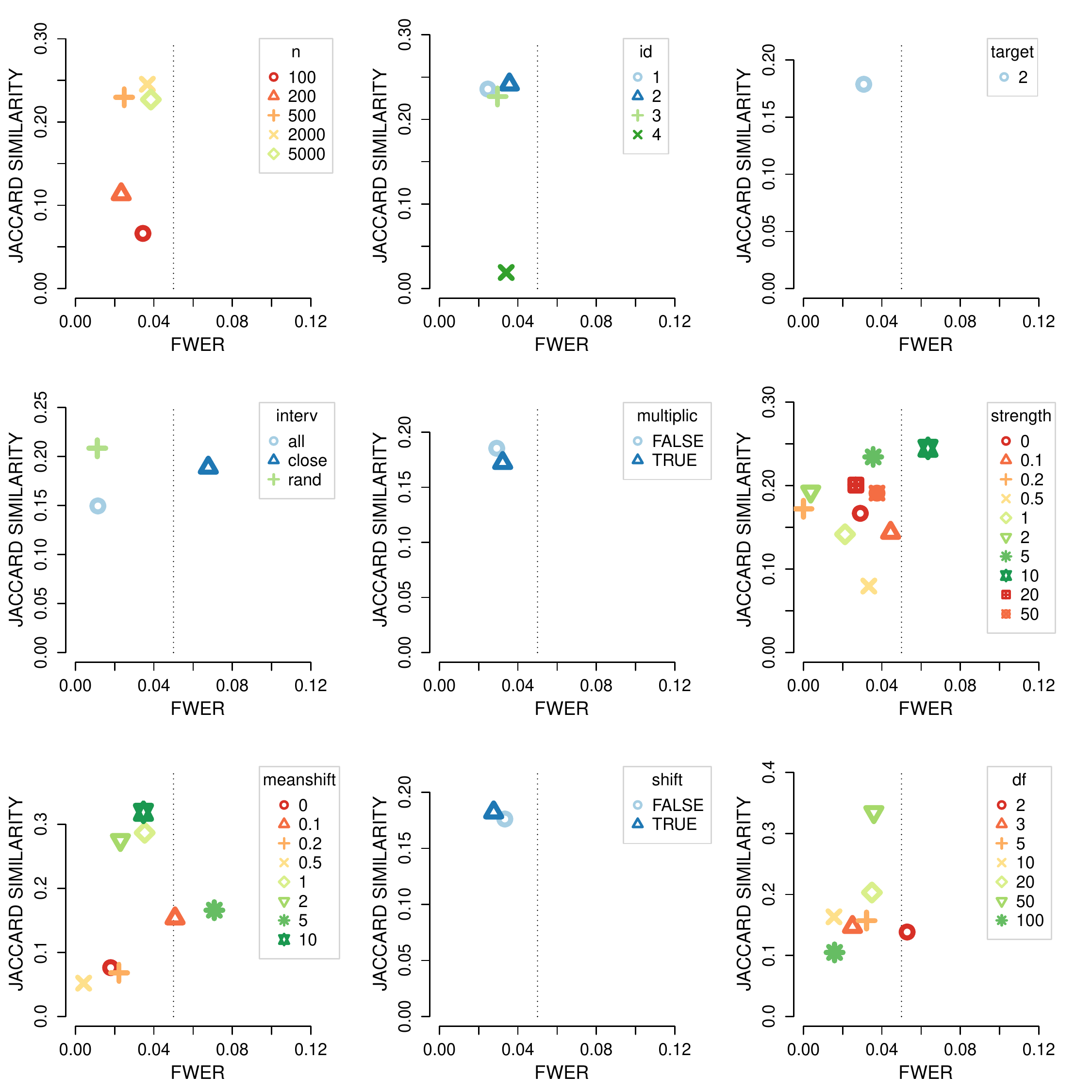}
\caption{\small Average Jaccard similarity over the conditional independence tests (A) -- (F) (y-axis) against average FWER (x-axis) when estimating the parents of variable 2. The figure is otherwise generated analogously to Figure~\ref{fig:performance_by_param}.}\label{fig:performance_by_param2}
\end{figure*}

\begin{figure*}[!t]
\center
    \includegraphics[page = 1, width=0.95\textwidth, keepaspectratio=true, trim = {0 0 0 0}, clip]{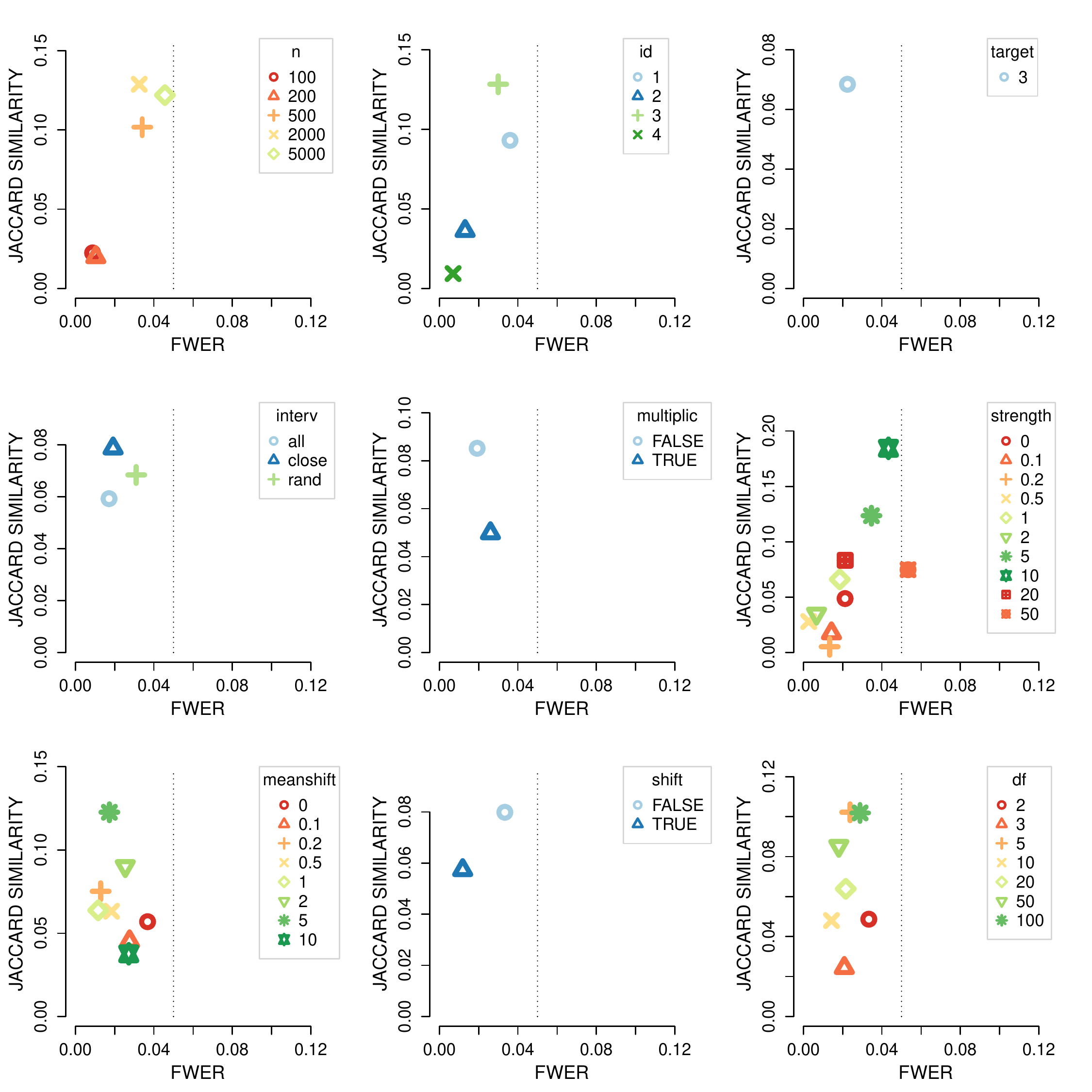}
\caption{\small Average Jaccard similarity over the conditional independence tests (A) -- (F) (y-axis) against average FWER (x-axis) when estimating the parents of variable 3. The figure is otherwise generated analogously to Figure~\ref{fig:performance_by_param}. }\label{fig:performance_by_param3}
\end{figure*}

\begin{figure*}[!t]
\center
    \includegraphics[page = 1, width=0.95\textwidth, keepaspectratio=true, trim = {0 0 0 0}, clip]{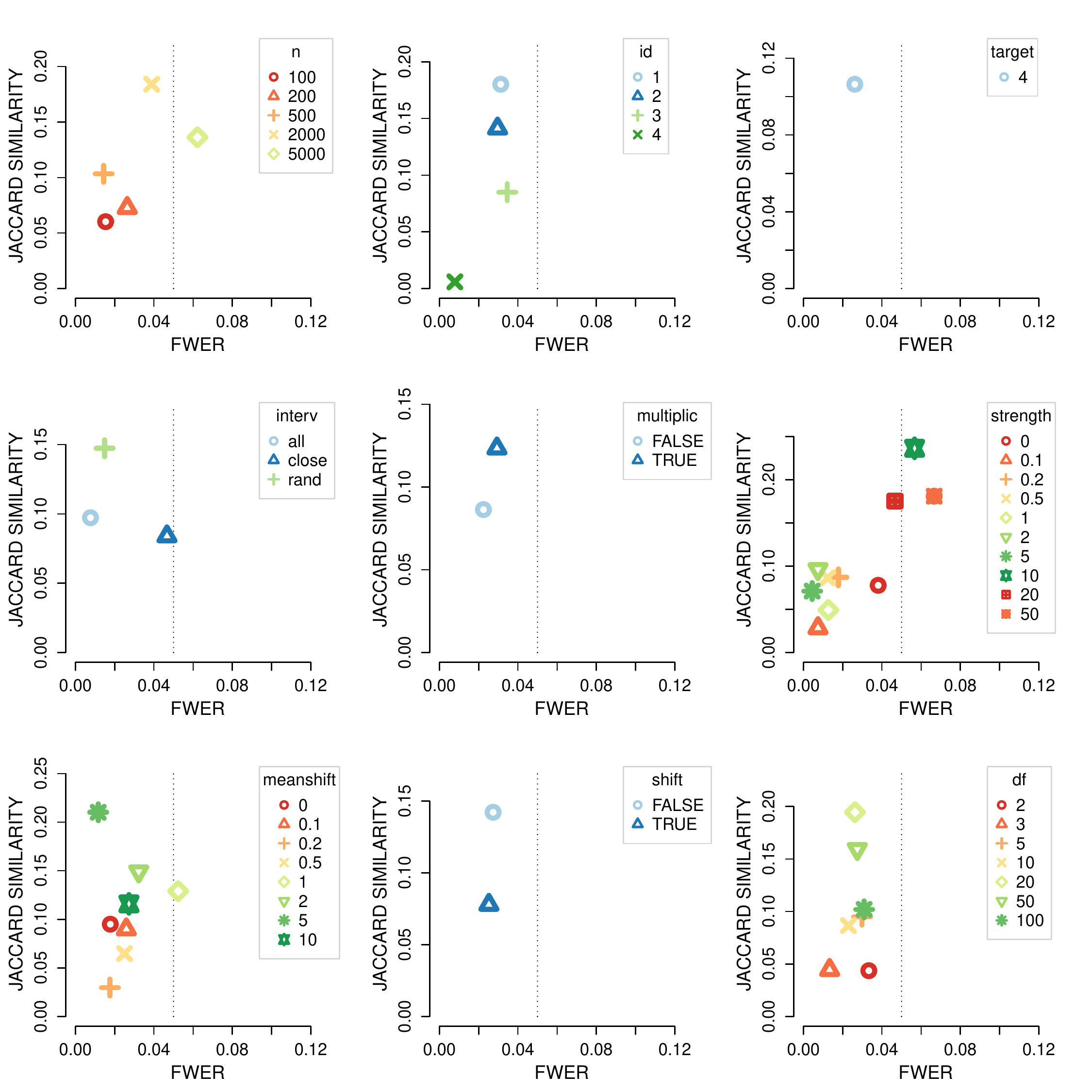}
\caption{\small Average Jaccard similarity over the conditional independence tests (A) -- (F) (y-axis) against average FWER (x-axis) when estimating the parents of variable 4. The figure is otherwise generated analogously to Figure~\ref{fig:performance_by_param}. }\label{fig:performance_by_param4}
\end{figure*}

\begin{figure*}[!t]
\center
    \includegraphics[page = 1, width=0.95\textwidth, keepaspectratio=true, trim = {0 0 0 0}, clip]{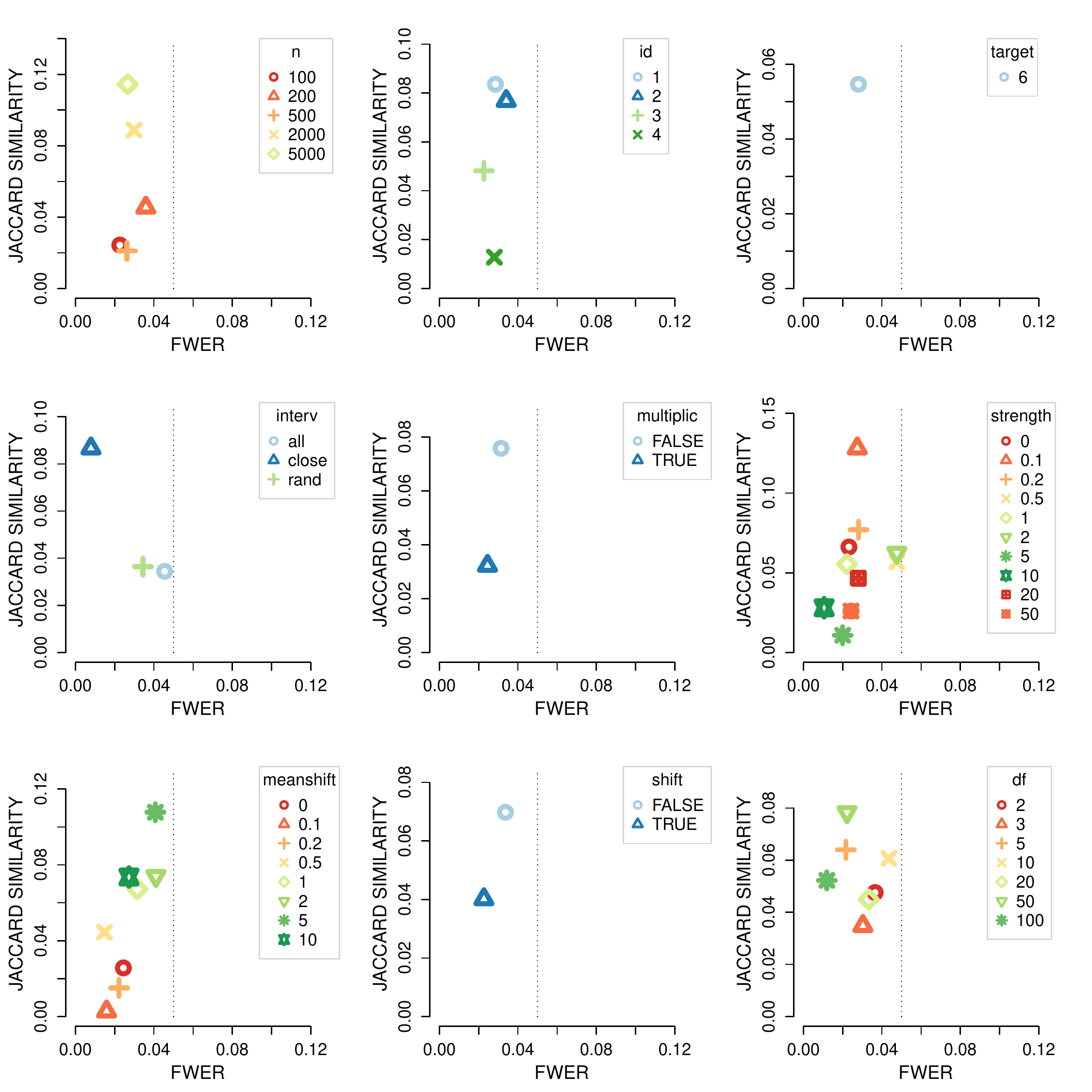}
\caption{\small Average Jaccard similarity over the conditional independence tests (A) -- (F) (y-axis) against average FWER (x-axis) when estimating the parents of variable 6. The figure is otherwise generated analogously to Figure~\ref{fig:performance_by_param}.}\label{fig:performance_by_param6}
\end{figure*}

 Figures~\ref{fig:performance_by_param2} -- \ref{fig:performance_by_param6} show the equivalent to Figure~\ref{fig:performance_by_param}, separately for target variables 2, 3, 4 and 6. 
For the sample size $n$, we observe that increasing it from 2000 to 5000 decreases power in case of target variable 4. This behavior can be explained by the fact that when testing $S^{*}$ in Eq.~\eqref{eq:H0S}, the null is rejected too often as the bias in the estimation performed as part of the conditional independence test yields deviations from the null that become significant with increasing sample size. 
For the nonlinearity, we find that the function $f_4(x) = \sin(2\pi x)$ is the most challenging one among the nonlinearities considered. It is associated with very low Jaccard similarity values for the target variables that do have parents. 
For the intervention type, it may seem surprising that `all' does not yield the largest power. A possible explanation is that intervening on all variables except for the target yields more similar intervention settings---the intervention targets do not differ between environments 2 and 3, even though the strength of the interventions is different. So more heterogeneity between the intervention environments, i.e.\ also having different intervention targets,  seems to improve performance in terms of Jaccard similarity.
Lastly, we see that power is often higher for additive parental contributions than for multiplicative ones. 

\FloatBarrier
\section{Example}\label{sec:example}
\begin{figure*}[!t]
\center
    \includegraphics[page = 10, width=1\textwidth, keepaspectratio=true, trim = {0 0 0 25}, clip]{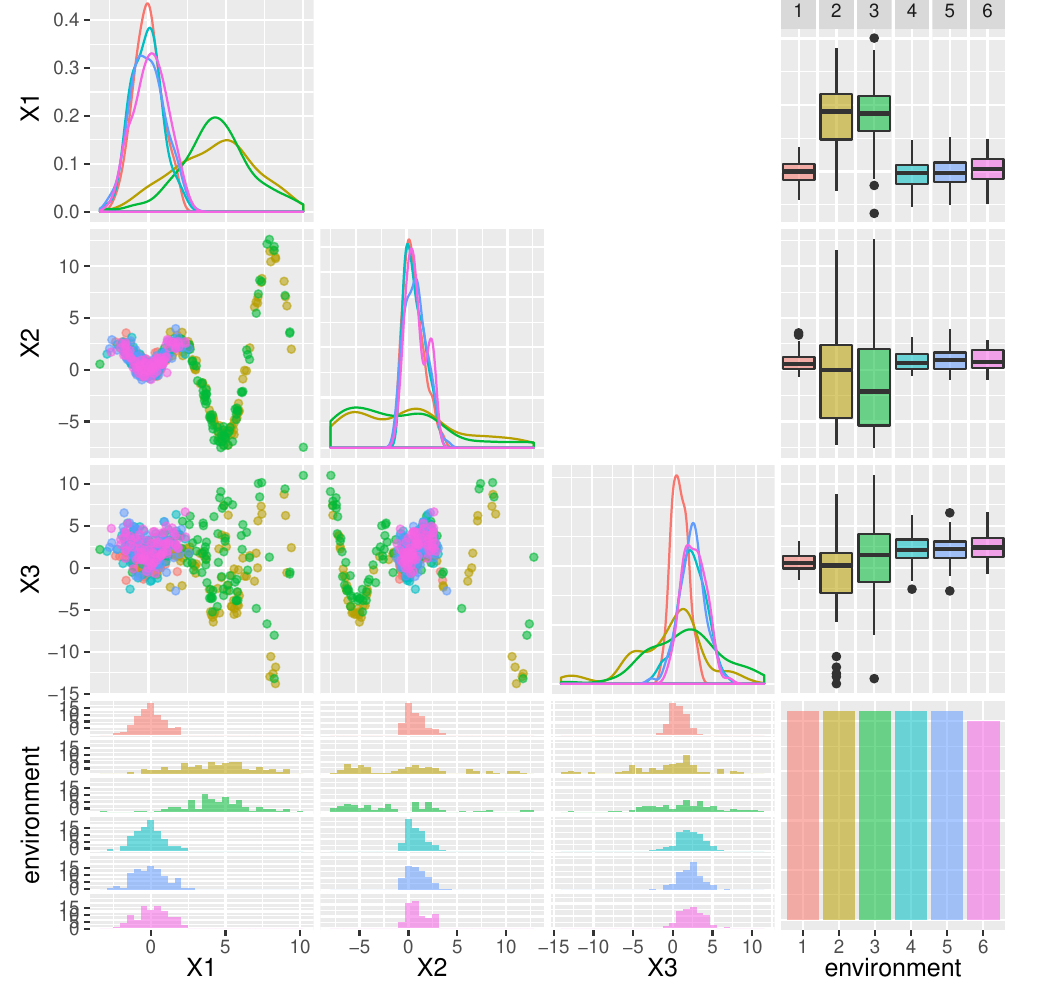}
\caption{\small Visualization of the sample considered in the example in Appendix~\ref{sec:example}.}\label{fig:example_sim}
\end{figure*}

Here we illustrate the methods presented in this manuscript by considering a causal DAG $X_1 \rightarrow X_2 \rightarrow X_3$. Figure~\ref{fig:example_sim} visualizes the generated data. There are six environments with shift interventions. The latter act on $X_1$ in two environments (green, yellow) and on $X_3$ in four environments (green, cyan, blue, magenta). The red environment consists of observational data. We run the proposed approaches (A) -- (F) to retrieve the parents of $X_2$, i.e.\ $S^* = \{X_1 \}$. Below we give an overview of which sets were accepted by the respective methods with $\alpha = 0.05$. We see that approaches (A), (B)(i)+(ii), (E)(i)-(iii) and (F) retrieve $S^*$ correctly, while the other approaches return the empty set. 

{\small
\begin{tabular}{p{20mm} p{25mm} p{25mm} p{25mm} p{25mm} p{25mm}}
CIT & $S_0 = \{\}$ & $S_1 = \{X_1\}$ & $S_2 = \{X_3\}$ & $S_3 = \{X_1, X_3\}$ & $\hat{S}$  \\
\hline
\abovespace\belowspace
(A)  &  & \checkmark & & \checkmark &  $\{X_1\}$ \\
(B)(i) &  & \checkmark & & \checkmark &  $\{X_1\}$ \\
(B)(ii) &  & \checkmark & & \checkmark &  $\{X_1\}$ \\
(B)(iii) & & & & &  $\{\}$  \\
(B)(iv) & & & & &  $\{\}$  \\
(C)  &  & \checkmark & \checkmark & \checkmark &  $\{\}$ \\
(D)(i)  & \checkmark & \checkmark & & \checkmark &  $\{\}$  \\
(D)(ii) & \checkmark & \checkmark & \checkmark & \checkmark & $ \{\} $ \\ 
(D)(iii) & & & & &  $\{\}$  \\
(D)(iv) &  \checkmark & & &  \checkmark &  $\{\}$  \\
(E)(i)  &  & \checkmark & & \checkmark &  $\{X_1\}$ \\
(E)(ii) &  & \checkmark & & \checkmark &  $\{X_1\}$ \\
(E)(iii)  &  & \checkmark & & & $\{X_1\}$ \\
(E)(iv) & & & & &  $\{\}$  \\
(F) &  & \checkmark & & \checkmark &  $\{X_1\}$ \\
\end{tabular}
}

\end{document}